\documentclass[12pt,notitlepage,a4paper]{article}
\pdfoutput=1
\usepackage[a4paper,margin=1in]{geometry}
\usepackage{amsmath,amssymb,amscd,amsfonts}
\usepackage{mathtools}
\numberwithin{equation}{section}
\usepackage{url}
\usepackage[shortlabels]{enumitem}
\usepackage{tikz} 
\usetikzlibrary{decorations}
\usetikzlibrary{arrows,decorations.markings}         
\usepackage[hidelinks]{hyperref}
\usepackage[open,
  openlevel=2,
  atend,
  numbered]{bookmark}
\usepackage{graphicx}
\usepackage[final]{pdfpages}
\usepackage{threeparttable}
\usepackage{caption}
\usepackage{float}
\usepackage{stackengine}
\usepackage{scalerel}
\usepackage[bf]{titlesec}
\titlespacing{\section}{3pc}{2pc}{0.8pc}
\titleformat{\section}
  {\centering\large\bfseries} 
  {\Large\thesection}
  {0.5em} 
  {\Large}  
\titlespacing{\subsection}{0pc}{2pc}{0.8pc}
\titleformat{\subsection}
  {\centering\bfseries} 
  {\normalsize\thesubsection}
  {0.5em} 
  {\normalsize}  
\usepackage{tocloft}

\newlength{\mylen}	
\urldef\footurlb\url{cocalc.com/dfriedan/DM/SM}
\urldef\footurla\url{physics.rutgers.edu/~friedan}
\def\eq{\begin{equation}}
\def\en{\end{equation}}
\def\eqg{\eq\begin{gathered}}
\def\eng{\end{gathered}\en}
\def\eqa{\eq\begin{aligned}}
\def\ena{\end{aligned}\en}
\def\Reals{\mathbb{R}}

\def\SO{\mathrm{SO}}
\def\SU{\mathrm{SU}}
\def\U{\mathrm{U}}
\def\Spin{\mathrm{Spin}}
\DeclareMathOperator{\cn}{cn}

\def\expval#1{\langle \, #1 \,\rangle}
\def\Vol{\mathrm{Vol}}
\def\gauge{\mathrm{gauge}}

\def\tr{\mathrm{tr}}

\def\EW{\scriptscriptstyle\mathrm{EW}}
\def\vH{v}
\def\kB{k_{\scriptscriptstyle \mathrm{B}}}

\def\phys{\scriptstyle \mathrm{phys}}
\def\grav{\scriptscriptstyle \mathrm{grav}}

\def\Higgs{\scriptscriptstyle \mathrm{Higgs}}
\def\Hubble{\scriptscriptstyle \mathrm{Hubble}}
\def\CGF{\scriptscriptstyle \mathrm{CGF}}
\def\CMB{\scriptscriptstyle \mathrm{CMB}}
\def\CDM{\scriptscriptstyle \mathrm{CDM}}

\def\ord{\mathrm{ordinary}}
\def\curvature{\text{curvature}}
\def\GeV{\text{\footnotesize GeV}}
\def\sunit{\text{\footnotesize s}}
\def\epsilonb{\epsilon}
\def\EhatCGF{\hat E_{\CGF}}
\def\ECGF{E_{\CGF}}
\def\ntimes{\,{\times}\,}

\def\gtwo{g}
\def\lambdaH{\lambda}

\stackMath
\def\dyhat{-0.2ex}
\newcommand\myhat[1]{\ThisStyle{%
              \stackon[\dyhat]{\SavedStyle#1}
                              {\SavedStyle\hat{\phantom{#1}}}}}
\def\that{\kern0.1em\myhat{\kern-0.1em t}}                              

\def\eff{{\scriptscriptstyle \mathrm{eff}}}
\def\ssm{{\scriptscriptstyle \mathrm{m}}}
\def\ssc{{\scriptscriptstyle \mathrm{c}}}

\def\ssLambda{{\scriptstyle \Lambda}}
\def\Kunit{\text{\small K}}
\def\CP{CP }
\def\kg{\mathrm{kg}}
\def\munit{\mathrm{m}}
\def\cmunit{\text{cm}}
\def\Msun{\textup{M}_{\odot}}
\def\texttilde{\raise-0.7ex\hbox{\!\texttt{\char`\~}}}
\begin{document}
\def\title{A theory of the dark matter}
\begin{center}
{\LARGE \title}
\vskip4ex
{\large Daniel Friedan}
\vskip2ex
{\it
New High Energy Theory Center
and Department of Physics and Astronomy,\\
Rutgers, The State University of New Jersey,\\
Piscataway, New Jersey 08854-8019 U.S.A. and
\vskip1ex
Science Institute, The University of Iceland,
Reykjavik, Iceland
\vskip1ex
\href{mailto:dfriedan@gmail.com}{dfriedan@gmail.com}
\qquad
\href{https://physics.rutgers.edu/\textasciitilde friedan/}
{physics.rutgers.edu/\texttilde friedan}
}
\vskip2ex
March 22, 2022
\end{center}

%
%
%
\begin{center}
\vskip3ex
{\sc Abstract}
\vskip2.5ex
\parbox{0.96\linewidth}{
\hspace*{1.5em}
In an earlier paper I proposed a highly symmetric semi-classical
initial condition to describe the universe in the period leading up to
the electroweak transition and completely determine all cosmology
after that.  Nothing beyond the Standard Model is assumed.  Inflation
is not needed.  The initial symmetry allows no adjustable parameters.
It is a complete theory of
the Standard Model cosmological epoch,
predictive and falsifiable.
Here, the time evolution of the initial condition is
calculated in the classical approximation.
The fields with nontrivial classical
values are the $\SU(2)$-weak gauge field (the cosmological gauge field
or CGF) and the Higgs field.  The CGF produces the electroweak
transition then evolves as a non-relativistic perfect fluid
($w_{\CGF}\approx 0$).  At the present time, i.e.~when $H=H_{0}$, the
CGF energy density satisfies $\Omega_{\Lambda}+\Omega_{\CGF}=1$.  The
CGF is the dark matter.  The dark matter is a classical phenomenon of 
the Standard Model. The classsical universe
contains only the dark matter, no ordinary matter.  At next to leading order the
fluctuations of the Standard Model fields will provide a calculable,
relatively small amount of ordinary matter such that
$\Omega_{\Lambda}+\Omega_{\CGF}+\Omega_{\ord}=1$.
}
\end{center}

\vspace*{-2ex}

%
%
%
\begin{center}
\tableofcontents
\end{center}
%
%
\section{Introduction}

I proposed in \cite{Friedan:2020poe} 
that the  initial condition 
of the universe leading up
to the electroweak transition
was a specific
highly symmetric semi-classical state of the Standard Model
and General Relativity.
The initial condition
is a classical SU(2)-weak cosmological gauge field (the CGF).
The  CGF cosmology has some attractive features:
\begin{itemize}
\item
Nothing beyond the Standard Model and General Relativity is assumed.

\item 
There are no adjustable parameters.

The initial symmetry and sufficiently large initial energy completely determine the initial state.
The initial condition is a specific state whose time evolution
describes all of the Standard Model cosmological epoch,
i.e.~from the electroweak transition onward ---
the period when the Standard Model and General 
Relativity govern physics.
\item
The state of the universe is semi-classical.
Its time evolution
can be calculated by a systematic
 expansion
around the classical CGF.

\item
It is a complete and calculable theory of cosmology,
predictive and falsifiable.

\item
It is an alternative to
inflation.
The initial symmetry implies spatial isotropy and homogeneity.
We will see here that a sufficiently large initial energy in the CGF
results in the observed  flatness of the present universe.

\end{itemize}
Here we calculate the time evolution of the initial condition
in the leading order, 
classical approximation, i.e.~ignoring the fluctuations of the fields.
\begin{itemize}
\item 
After the CGF produces the electroweak transition, it drives the expanding 
universe
as a non-relativistic fluid ($w_{\CGF} = 0$)
to become the dark matter in the present universe.
The dark matter is a classical phenomenon in the Standard Model.
\item
The universe contains only dark matter
in the leading order, classical approximation.
The ordinary matter appears at the next order
as a relatively small perturbation of the dark matter 
universe by the fluctuations of the Standard Model fields.
\item
We derive the dark matter equation of state in analytic form.
\end{itemize}
In \cite{Friedan2022:DMStars} the 
Tolman-Oppenheimer-Volkoff stellar structure equations
for dark matter stars made of the CGF are solved numerically using
the CGF equation of state derived here.
\begin{itemize}
\item 
The possible CGF dark matter stars ---
the solutions of the TOV equations ---
have mass $M$ ranging from  $0$ up to $9.14 \ntimes 10^{-6}\,\Msun$
with radius $R$ taking certain specific values
between $5.23\, \cmunit$ and $13.6\,\cmunit$.
Presumably the  CGF has collapsed gravitationally into 
an ensemble of such dark matter stars.
For $M>5.09 \ntimes 10^{-6}\,\Msun$
multiple values of $R$ are possible,
suggesting the possibility of transitions
that would release gravitational energy on the order of $10^{41}\,\mathrm{J}$
in times on the order of $10^{-10}\,\mathrm{s}$.

\end{itemize}
In \cite{Friedan2022:Stability} it is shown that
\begin{itemize}
\item 
The initial condition is 
thermodynamically stable under fluctuations of the SU(2)-weak gauge 
field, a first step towards showing that the initial condition is physically natural.
\end{itemize}
The classical CGF is the skeleton of cosmology --- a universe 
containing only dark matter.
The fluctuations of the Standard 
Model fields will flesh out the skeleton with ordinary matter.
The skeleton has the right form.
It remains to calculate
the higher order corrections
to check whether the CGF cosmology matches the observed universe.

\subsection{Assumptions}
\noindent
The only assumptions are:
\begin{enumerate}
\item The universe is governed by the Standard Model
and General Relativity (with cosmological constant).
Nothing beyond the Standard Model is assumed.
Nothing beyond the known laws of physics is assumed.
\item The universe is a 3-sphere.
\item The state of the universe is 
invariant under a $\Spin(4)$ symmetry group
that acts on the 3-sphere as $\SO(4)$
and on the Standard Model fields
such that the $\SU(2)$-weak doublets transform as spinors.
\item The initial energy in the Standard Model fields is $ >  10^{107}$ in natural units.
\end{enumerate}

\subsection{The CGF cosmology}

These assumptions result in a unique cosmology.
There are no adjustable 
parameters.
The cosmology is semi-classical.
The leading order approximation --- the classical approximation --- 
is solvable in terms of elliptic functions and elliptic 
integrals.
The universe at leading order
contains only the CGF, no ordinary matter.
The ordinary matter appears at next to leading order as a perturbation 
due to the fluctuations of the 
Standard Model fields around the classical solution.

\begin{enumerate}[leftmargin=*]
\item
The $\Spin(4)$ symmetry and high initial energy imply that the state 
of the universe is semi-classical --- a solution of the classical 
equations of motion plus small fluctuations.
The fluctuations are ignored in the leading order, classical approximation.

\item 
The only nontrivial fields in the $\Spin(4)$-symmetric classical 
solution are
\begin{itemize}
\item
the homogeneous, isotropic space-time metric characterized by the 
radius $R(\that)$ of the 3-sphere universe as a function
of conformal time $\that$,
\item
the Higgs
field $\phi$ which is fixed by the $\Spin(4)$ symmetry at the local maximum $\phi=0$ of the Higgs
potential,
\item
a $\Spin(4)$-symmetric $\SU(2)$-weak gauge field (the CGF).
\end{itemize}

\item
The $\Spin(4)$-symmetric $\SU(2)$-weak gauge field 
is described by a single degree of freedom $\hat b(\that)$.
The equation of motion for $\hat b(\that)$ is solved 
analytically by an elliptic function.
$\hat b(\that)$ oscillates anharmonically
with dimensionless energy $\EhatCGF$ in the natural energy unit
$\hbar/R(\that)$ (with $c=1$).
It will turn out that 
 $\EhatCGF >10^{107}$ is needed for
the electroweak transition to take place and 
for the present curvature of the universe to be consistent with the 
observed flatness.
\item
There is a natural parametrization of the scale of the universe,
$a(\that) =\epsilon R(\that)$, where $\epsilon = (8\EhatCGF)^{-1/4}$.
The local physics in terms of $a(\that) $ is independent 
of the particular value of the energy $\EhatCGF>10^{107}$
so the CGF cosmology has no adjustable parameters.

\item
The universe expands,
driven by the gauge field energy and the energy in the Higgs field at 
$\phi=0$.
The CGF behaves effectively as a perfect fluid, a mixture of radiation
with a small amount of vacuum energy,  $w_{\CGF}\approx 1/3$.

\item
The CGF oscillation is much faster than the expansion
so an adiabatic approximation is justified,
averaging over the CGF oscillation.
The rapidly oscillating CGF contributes an 
effective $\phi^{\dagger}\phi$ term to the Higgs potential which 
stabilizes
$\phi$ at $0$.  The effective $\phi^{\dagger}\phi$ 
term is proportional to $1/a^{2}$ so the stabilizing effect weakens as the universe 
expands.

\item
The stabilizing effect becomes insufficient
when the scale reaches $a_{\EW}= 0.585\, \hbar/m_{\Higgs}$.
The electroweak transition starts.
The Higgs field begins to move away from $0$ towards its vacuum 
expectation value.

\item
The elliptic function $\hat b(\that)$ is  periodic in both imaginary 
time and real time.  The imaginary time period defines an inverse temperature.
The fluctuations of the fields are initially in 
the specific thermal state defined by this periodicity in imaginary time.

The initial condition leading up to $a_{\EW}$ is a specific 
semi-classical state of the Standard Model and General Relativity.
All of cosmology after $a_{\EW}$ is calculable from first principles
starting from this initial condition.

\item
The adiabatic approximation continues to be valid after $a_{\EW}$.
The adiabatic equations of motion for the CGF after $a_{\EW}$ are solved analytically by
elliptic functions and elliptic integrals.

\item
After $a_{\EW}$,
over the next 1 or 2 ten-folds of expansion,  $\phi^{\dagger}\phi$ tracks 
the minimum of the evolving effective Higgs potential to the vacuum 
expectation value $v^{2}/2$.
The CGF evolves adiabatically to become
harmonic oscillation at the bottom of the massive gauge field potential.

\item
From about $10^{2}\,a_{\EW}$ onward
the expansion of the universe is
driven by the CGF 
energy density $\rho_{\CGF}= 0.890\, m_{\Higgs}/a^{3}$.
From about $10^{2}\,a_{\EW}$ onward
the CGF is a non-relativistic perfect fluid
--- a perfect fluid with equation of state parameter $w_{\CGF}= 
0$.

Figure~\ref{fig:SMepoch} shows the evolution of $w_{\CGF}$ and 
$\phi^{\dagger}\phi$ with the scale of the universe.
\begin{figure}
\begin{center}
\captionsetup{justification=centering,margin=0.12\linewidth}
\includegraphics[scale=0.8]{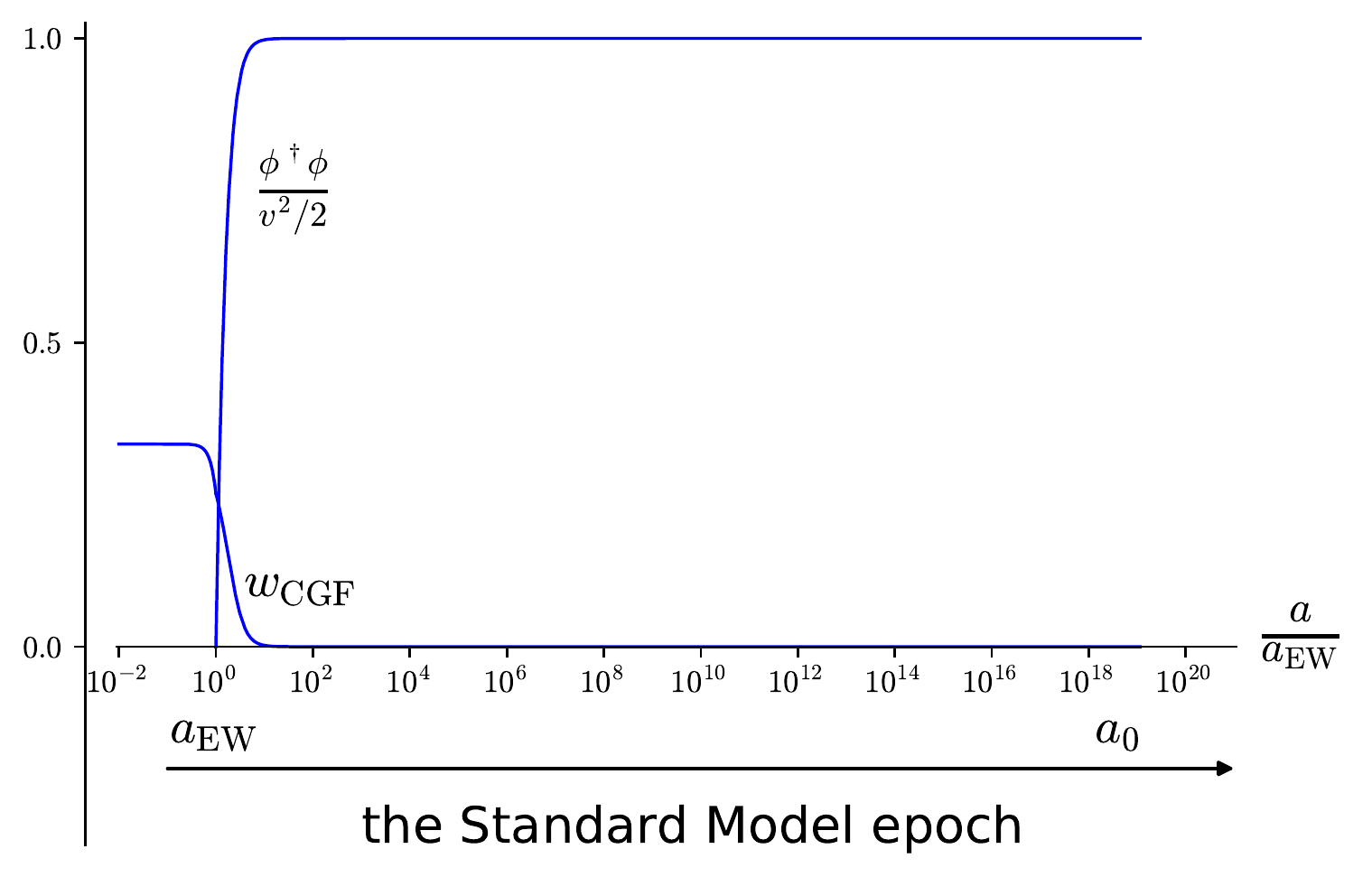}
\caption{
\label{fig:SMepoch}
The CGF is a radiation fluid at high density, before $a_{\EW}$.
At low density, after about $10^{2}a_{\EW}$, it is 
a non-relativistic fluid.
}
\end{center}
\end{figure}

\item
The rate of expansion $H$ decreases with time,
reaching the Hubble constant $H_{0}$
at $a_{0}= 1.40  \ntimes 10^{-8}\,\sunit$.
The condition
$H=H_{0}$ identifies the present.

\item
The present curvature is
\eq
\label{eq:presentflatness}
\frac1{H_{0}^{2} R_{0}^{2}} = \frac{1}{H_{0}^{2} a_{0}^{2}(8 \hat 
E_{\CGF})^{1/2}} 
\en
The observed flatness of the universe
puts a lower bound on the initial energy.
\eq
\frac1{H_{0}^{2} R_{0}^{2}} < 0.001
\quad\Longleftrightarrow\quad
\EhatCGF > 10^{107}
\en

\item 
In the leading, classical approximation
the only matter in the universe is
the CGF.
The present curvature is negligible
so the present normalized energy density is
\eq
\Omega = \Omega_{\Lambda}  + \Omega_{\CGF} = 1
\en
There is only  dark matter, the CGF.

\item
The fluctuations will provide the ordinary mattter.
The total energy density in the
semi-classical state at $H=H_{0}$ will be
$\Omega_{\ssLambda}+ \Omega_{\CGF}+ \Omega_{\ord}  =1$.
The ratio $\Omega_{\ord}/\Omega_{\CGF}$ will have to
agree with the observed
ratio $\Omega_{\ord}/\Omega_{\CDM}$.
\end{enumerate}

There are no adjustable parameters in the theory.
The energy $\EhatCGF$, the one apparently adjustable parameter,
is irrelevant to the local physics.
Everything in the theory is calculable.
Every calculation of an observable
quantity is an 
opportunity to confirm or refute the theory.
If the theory works, 
it will provide a first principles cosmology
of the Standard Model epoch.
And it will be a sharp target for 
theories of cosmology before the electroweak transition.
A theory of the early universe
will be successful if and only if
it produces
the Standard Model plus General Relativity in
a spherical universe with
$\Spin(4)$ symmetry and
high initial energy.

\subsection{Physical parameters}

Table~\ref{table:physicalparameters} shows the physical parameters 
(in $c=1$ units) that are used in the calculations.
The values are taken from \cite{PDG2021}.
$\kappa = 8\pi G$ is the gravitational constant.
$m_{\Higgs}$ is the mass of the Higgs boson.
$H_{0}$ is the Hubble constant.
${}g$ is the $\SU(2)$ gauge coupling 
constant.  
$\lambda$ is the Higgs coupling constant.
$\Omega_{\CDM}=\rho_{\CDM}/\rho_{c}$ is the present dark matter density normalized to the 
critical density $\rho_{c} = 3H_{0}^{2}/\kappa$.  $\Omega_{\ord}$ is the normalized 
density of ordinary matter.  $\Omega_{\Lambda}$ is the normalized 
dark energy density (assumed due to the cosmological constant).
The normalized curvature is $-\Omega_{\curvature}$.  
The normalization is such that
$\Omega_{\Lambda} +\Omega_{\CDM}+\Omega_{\ord}+\Omega_{\curvature}= 1$.
\begin{table}[H]
$$
{ \everymath={\displaystyle}
\def\arraystretch{2.5}
\begin{array}{|@{}rc@{\!}c@{\,}l@{\quad}|@{\quad}c@{\,}l|}
\hline
&t_{\grav}= &\, (\hbar\kappa )^{\frac12} & =2.70 \ntimes 
10^{-43}\,\sunit&
g^{2} =  0.426 \qquad \lambda^{2} = 0.258&
\\[1ex]
\hline
&t_{\Higgs}= &\, \frac{\hbar}{m_{\Higgs}} & = 5.26 \ntimes 10^{-27}\,\sunit&
\Omega_{\CDM} = 0.266 \quad \Omega_{\ord} =0.049&
\\[1ex]
\hline
&t_{\Hubble}=  &\, \frac1{H_{0}} &  = 4.58 \ntimes 10^{17}\,\sunit&
\Omega_{\ssLambda} = 0.685
\quad |\Omega_{\curvature}| < 0.001 &
\\[1ex]
\hline
\end{array}
\def\arraystretch{1}
}
$$
\caption{
\label{table:physicalparameters}
Physical parameters (in $c=1$ units)
}
\end{table}

The calculations in the paper are in the leading order, classical 
approximation,
subject to higher order corrections.
The results are shown to three decimal places
to indicate that
experimental accuracy will be the only limitation
once the higher order corrections are calculated.

\subsection{Organization of the paper}

The paper is organized as follows.
Section 2 summarizes \cite{Friedan:2020poe}, describing the $\Spin(4)$-symmetric 
initial condition, 
covering points 1 through 8 above.
Section 3 cites some mathematical literature.
Section 4 calculates the adiabatic time evolution of the CGF after 
the start of the electroweak transition at $a_{\EW}$.
Section 5 identifies the present time by the condition $H=H_{0}$,
expresses the present curvature in terms
of the CGF initial energy,
and shows that the present energy density in the CGF gives the observed 
flatness,
$ \Omega_{\CGF}+\Omega_{\Lambda}  -1 < 0.001$.
The CGF equation of state calculated in Sections 2 and 4 is 
summarized.
Section 6 collects comments and questions.
Appendix~\ref{app:elliptic} lists
some identities for the elliptic functions and elliptic integrals.

The numerical calculations are done in SageMath \cite{sagemath9.4}
using the mpmath arbitrary-precision floating-point arithmetic 
library \cite{mpmath}.
The Sagemath notebooks along with printouts
are provided in the Supplemental Material \cite{Friedan2022:AtheorySuppMat}.

\section{Spin(4)-symmetric initial condition}

This section is a summary of \cite{Friedan:2020poe} with some small changes of 
notation.
Details of the calculations are shown in the
Supplemental Material of \cite{Friedan:2020poe}.

\subsection{Spin(4)-symmetric physics on the 3-sphere}

Let space be a 3-sphere.
Write $S^{3}$ for the unit 3-sphere in $\Reals^{4}$ with metric
$\hat g_{ij}(\hat x)$.
\eq
\hat x \in  \Reals^{4}
\qquad
(\hat x^{\mu}) = (\hat x^{i},\hat x^{4}) 
\qquad
\delta_{\mu\nu} \hat x^{\mu} \hat x^{\nu} = 1
\qquad
\hat g_{ij}(\hat x)d\hat x^{i}d\hat x^{j} = \delta_{\mu\nu} d\hat x^{\mu} d\hat x^{\nu}
\en
Identify $S^{3}$  with $\SU(2)$ by
\eq
\hat x \in S^{3}
\:\:\:\longleftrightarrow\:\:\:
g_{\hat x} = \hat x^{4} \mathbf1 + \hat x^{k}  i^{-1} \sigma_{k} \;\in\SU(2)
\en
$\sigma_{k}$ being the Pauli matrices.
$\Spin(4)$ is $\SU(2)_{L}{\ntimes} \SU(2)_{R}$.
It acts on $S^{3}$ as $\SO(4)$.
\eq
U=(g_{L},g_{R})
\qquad
g_{U\hat x} = g_{L} \,g_{\hat x}\, g_{R}^{-1}
\en
The general $\SO(4)$-symmetric space-time metric is
\eq
ds^{2} = R(\that)^{2}\left(
-d\that{\;\!}^{2} +  \hat g_{ij}(\hat x)d\hat x^{i}d\hat x^{j}
\right)
\en
$\that$ is conformal cosmological time.
$R(\that)$ is the radius of the spatial 
3-sphere at conformal time $\that$.
Co-moving time $t$ is given by
$dt = R(\that) d \that$.

Let $\Spin(4)$ act on
the $\SU(3)$ and $\U(1)$ gauge bundles of the Standard Model as
product bundles, so the only $\Spin(4)$-symmetric gauge fields 
are the trivial gauge fields,
$A_{\mu}^{\SU(3)}=A_{\mu}^{\U(1)}=0$.
Let the $\SU(2)$ gauge bundle be identified with the spinor bundle over 
$S^{3}$.  
An $\SU(2)$ doublet field such as the Higgs field $\phi$ transforms 
under $\Spin(4)$ by
\eq
U=(g_{L},g_{R})
\qquad
\phi \mapsto U\phi
\qquad
U \phi(\that,  \hat x) =  g_{L} \,\phi(\that ,U^{-1} \hat x)
\en
The only $\Spin(4)$-symmetric Higgs field is $\phi=0$.
There is a one parameter family of $\Spin(4)$-symmetric $\SU(2)$ gauge fields
on $S^{3}$ with covariant derivative
\eqg
D^{\SU(2)}_{i} = \hat\nabla_{i} + \hat b \, \hat \gamma_{i}(\hat x)
\eng
where the $\hat \gamma_{i}(\hat x)$ are the Dirac matrices at $\hat x$
and $\hat \nabla$ is the metric covariant derivative, 
\eqg
\hat \gamma_{i}(\hat x)
=  \frac12 g_{\hat x} \hat \partial_{i} (g_{\hat x}^{-1})
\qquad
(\hat \gamma_{i} \hat \gamma_{j} + \hat \gamma_{j}\hat \gamma_{i})(\hat x)
= -\frac12 \hat g_{ij}(\hat x)
\\[1ex]
\hat \nabla_{i} = \hat \partial_{i} + \hat \gamma_{i}
\qquad
\hat \partial_{i} = \frac{\partial}{\partial \hat x^{i}}
\qquad
\hat \nabla_{i}\hat \gamma_{j} = 0
\eng
In space-time
the general $\Spin(4)$-symmetric $SU(2)$ gauge field
in unitary gauge is
\eqg
D^{\SU(2)}_{0} = 0
\qquad
D^{\SU(2)}_{i} = \hat \nabla_{i} + \hat b(\that) \hat \gamma_{i}(\hat x)
\eng
The cosmological gauge field (the CGF) is
$B^{\CGF}_{i} = \hat b(\that)\, \hat \gamma_{i}(\hat x)$
or, for short, $\hat b(\that)$.

The general $\Spin(4)$-symmetric classical state is
characterized by the pair of functions
$R(\that)$ and $\hat b(\that)$ subject to the classical equations of motion.
The Yang-Mills action (as written in \cite{PDG2021}) is
\eq
\frac1\hbar S_{\gauge} =\int
\frac1{2 \gtwo ^{2}}
\tr(- F_{\mu\nu} F^{\mu\nu})
\, \sqrt{-g} \,d^{4}x
\en
The action of the $\Spin(4)$-symmetric gauge field is
\eq
\label{eq:bhataction}
\frac1\hbar S_{\gauge} =
\Vol(S^{3})\frac{3  }{g^{2}}
\int
\bigg[-\frac12 \bigg(\frac{d\hat b}{d\that}\bigg)^{2}+ \frac12 
(\hat b^{2}-1)^{2}\bigg]
d\that
\qquad
\Vol(S^{3}) = 2\pi^{2}
\en
The action is
independent of $R$ because the Yang-Mills theory is
conformally invariant.
$\hat b$ is an anharmonic oscillator
with equation of motion
\eq
\frac{d^{2}\hat b}{d\that^{2}} + 2 \hat b (\hat b^{2}-1) = 0
\en
The dimensionless energy
\eq
\EhatCGF = \frac12 \left(\frac{d\hat b}{d\that}\right)^{2} + \frac12(\hat b^{2}-1)^{2}
\en
is conserved.  
In co-moving time the action is
\eq
\label{eq:baction}
S_{\gauge} =
\frac\hbar{R}
\Vol(S^{3})\frac{3  }{g^{2}}
\int
\bigg[-\frac12 \bigg(\frac{d\hat b}{d t}\bigg)^{2}+ \frac12 
(\hat b^{2}-1)^{2}\bigg]
dt
\en
The physical energy is dual to co-moving time so
\eq
\label{eq:CGFenergy}
E^{\phys}_{\CGF} =  \frac{\hbar}{R}\Vol(S^{3})\frac{3 }{g^{2}} \EhatCGF
\en
The general classical solution (for $\EhatCGF>1/2$)
is the elliptic function
\eqg
\label{eq:classicalsoln}
\hat b(\that) = \frac{b(z)}{\epsilonb } 
\qquad
b(z) = k\cn( z,k)
\\[1ex]
z = \frac{\that}{\epsilonb } 
\qquad
k^{2}= \frac12 + \epsilonb ^{2}
\qquad
\EhatCGF = \frac1{8\epsilonb ^{4}}
\eng
Equation (\ref{eq:kcneqn}) is the relevant identity on the Jacobi 
elliptic function $\cn(z,k)$.
The CGF $\hat b(\that)$ oscillates between $\pm k/\epsilonb$ in the quartic potential 
of (\ref{eq:bhataction})
with period $\Delta \that = 4K\epsilon$
where $K=K(k)$ is the complete elliptic integral of the first kind.

\subsection{Initial CGF energy}

$\EhatCGF$ is the only adjustable parameter in the classical solution.
We start out assuming $\EhatCGF \gg 1$  
so that the CGF is semi-classical.
Later we will find that $\EhatCGF$ must be very large
if the cosmology is to be physically realistic.
We first find  $\EhatCGF > 10^{67}$ is needed
in order that the CGF produces the electroweak transition 
in an expanding universe.
Then we derive a formula for the present curvature of the universe,
\eq
-\Omega_{\curvature} = \frac{1}{H_{0}^{2} R_{0}^{2}}
= 1.07\ntimes 10^{51} \; \epsilon^{2} 
\en
The observed flatness of the 
present universe, $|\Omega_{\curvature}|< 0.001$, implies
\eq
\label{eq:energyepsilonbounds}
\epsilon 
 <  10^{-27}
\qquad
\EhatCGF >  10^{107}
\en
The bound (\ref{eq:energyepsilonbounds}) on $\EhatCGF$ 
replaces a handwaving argument in \cite{Friedan:2020poe} that gave a similar bound.
When $\EhatCGF$ is very large, $\epsilonb$ is very small, so to very high accuracy
\eqg
k^{2} = \frac12
\qquad
K = K(1/\sqrt2) =\frac{\Gamma(1/4)^{2}}{4 \pi^{1/2}} = 
1.854075\ldots
\eng

\subsection{Coordinates for local physics}
Scale the spatial coordinates $x = {\hat x}/{\epsilonb}$  to match 
the scaling of time $z =\that/\epsilonb$.\footnote{Note that the complex conformal time variable 
$z$ is \emph{not} the redshift.}
The space-time metric in the scaled coordinate system is
\eq
ds^{2} = a(z)^{2}\left(
-dz^{2} +   g_{ij}( x)d x^{i}d x^{j}
\right)
\qquad
a(z) = \epsilonb R(\that)
\en
The geometry of the scaled 3-sphere is
\eqg
x \in \frac1{\epsilonb } S^{3}
\qquad
x = \frac{\hat x}{\epsilonb} 
\qquad
\delta_{\mu\nu} x^{\mu}  x^{\nu} = \frac1{\epsilonb^{2}}
\qquad
g_{ij}(x)dx^{i} dx^{j}= \delta_{\mu\nu} d x^{\mu} d x^{\nu}
\\[1ex]
g_{ij}(x)dx^{i} dx^{j}= \frac1{\epsilonb^{2}} \hat g_{ij}(\hat x) d\hat x^{i} d\hat x^{j}
\qquad
g_{ij}(x) = \hat g_{ij}(\hat x)
\\[1ex]
\gamma_{i}(x) dx^{i} = \frac1{\epsilonb} \hat \gamma_{i}(\hat x) d\hat x^{i}
\qquad
\gamma_{i}(x) = \hat \gamma_{i}(\hat x)
\qquad
\gamma_{i} \gamma_{j} + \gamma_{j} \gamma_{i}  = -\frac12 g_{ij}
\eng
The CGF covariant derivative is
\eq
\label{eq:CGFcovariantderivative}
D_{i} = \nabla_{i}  +  b(z) \gamma_{i}(x)
\qquad
\nabla_{i} = \partial_{i}+ \epsilonb \gamma_{i}
\qquad
b(z)  = k \cn(z,k) 
\en
The $\Spin(4)$-symmetric gauge field action (\ref{eq:bhataction}) becomes
\eq
\label{eq:gaugeaction}
\frac1{\hbar} S_{\gauge} = \frac{\Vol(S^{3})}{\epsilonb^{3}}\frac{3  }{g^{2}}
\int
\bigg[-\frac12 \bigg(\frac{d b}{d z}\bigg)^{2}+ \frac12 
\left(b^{2}-\epsilonb^{2}\right)^{2}\bigg]
dz
\en
The local physics in the scaled coordinate system
is independent of $\epsilonb$
when $\epsilonb$ is very small.
Only global properties of the universe can depend on $\epsilonb$.

\subsection{Start of the electroweak transition at $a=a_{\EW}$}
The action of the Higgs field (as written in \cite{PDG2021}) is
\eqg
\label{eq:Higgsaction}
\frac1\hbar S_{\Higgs} =\int
\bigg [
 a^{-2} D_{\mu}\phi^{\dagger} D^{\mu}\phi
+  \frac12\lambdaH^{2}\bigg(\phi^{\dagger}\phi 
-  \frac{\vH ^{2}}2 \bigg)^{2}
\,\bigg]\, a^{4}\sqrt{-g} \,d^{4}x
\eng
where $x^{0} =z$, $D_{0}=\partial_{0}$, and $D_{i}$
is the $\SU(2)$-weak gauge covariant derivative.
The Higgs mass is $m_{\Higgs}=\hbar \lambda v$.
In the presence of the CGF
the covariant derivative
(\ref{eq:CGFcovariantderivative}) gives
\eq
 D_{\mu}\phi^{\dagger} D^{\mu}\phi
= \partial_{\mu}\phi^{\dagger} \partial^{\mu}\phi
+ \frac12 b(z) (\partial^{i}\phi^{\dagger} \gamma_{i}\phi
- \phi^{\dagger}\gamma_{i} \partial^{i}\phi)
+ \frac3{4} b(z)^{2} \phi^{\dagger}\phi
\en
with $\partial_{i}$ replacing $\nabla_{i}=\partial_{i}+\epsilonb 
\gamma_{i}$ because $\epsilonb$ is 
very small.

We make two assumptions to be verified later.
We assume that $a$ and $\phi$ change very slowly
compared to the CGF oscillation.
And we assume that the energy in the CGF and in the Higgs field at $\phi=0$ drives an 
expanding universe.

Given the that $a$ and $\phi$ are changing slowly,
$b(z)$ and $b(z)^{2}$ can be replaced in the action 
(\ref{eq:Higgsaction}) by their expectation values over a
period of oscillation.
\eq
\expval{b} = 0
\qquad
\expval{b^{2}} =
\frac1{4K}\int_{0}^{4K} k^{2} \cn^{2}(z,k) \; dz = \frac{\pi}{4 K^{2}}
\en
Equation (\ref{eq:cnintegrals}) is the relevant identity.
The effective action for $\phi$ is
\eqa
\label{eq:VHiggseff}
\frac1\hbar S_{\Higgs}^{\eff} &=\int
\left [
 a^{-2} \partial_{\mu}\phi^{\dagger} \partial^{\mu}\phi
+ V^{\eff}_{\Higgs}(\phi)
\,\right]\, a^{4}\sqrt{-g} \,d^{4}x
\\[2ex]
 V^{\eff}_{\Higgs}(\phi) 
&= \frac{\lambdaH^{2}\vH^{4}}8
+ \left(
\frac3{4}  \frac{\expval{b^{2}} }{a^{2}}
-\frac{\lambdaH^{2}\vH ^{2}}2 
\right)
\phi^{\dagger}\phi
+ \frac{\lambdaH^{2}}2  (\phi^{\dagger}\phi)^{2}
\ena
When $a(z)$ is small
the coefficient of $\phi^{\dagger}\phi$ is positive
so  $\phi=0$ is stable.  When the coefficient of $\phi^{\dagger}\phi$ becomes
negative, $\phi=0$ becomes unstable.  
The electroweak transition starts when
the quadratic term vanishes,
at $a = a_{\EW}$ given by
\eq
\label{eq:a_EW}
a_{\EW}
= \left(\frac{3 }{2} \frac{\expval{b^{2}}}{\lambdaH^{2}\vH^{2} }\right)^{\frac12}
=  \frac{(6\pi)^{\frac12}}{4 K} t_{\Higgs}
= 0.5854 \, t_{\Higgs}
= 3.08\ntimes 10^{-27}\,\sunit
\en

\subsection{CGF density and pressure for $a\le a_{\EW}$}

The energy-momentum tensors (derived in \cite{Friedan:2020poe}) are
\eqa
\label{eq:Tmunu}
\frac1{\hbar} T^{\gauge}_{\mu\nu}  &=
{\gtwo ^{2}}
\tr
\left (
- 2F_{\mu\sigma} F_{\nu}{}^{\sigma}
+ \frac12 g_{\mu\nu} F_{\rho\sigma} F^{\rho\sigma}
\right )
\\[1ex]
\frac1\hbar  T^{\phi}_{\mu\nu}
 &=
 2 D_{\mu}\phi^{\dagger} D_{\nu}\phi
-g_{\mu\nu}  D_{\sigma}\phi^{\dagger} D^{\sigma}\phi
-  g_{\mu\nu} \frac12 \lambdaH^{2} \left(\phi^{\dagger}\phi 
-  \frac{1}2 \vH ^{2} \right)^{2}
\ena
For the $\Spin(4)$-symmetric state in the classical, adiabatic 
approximation the total energy-momentum tensor is expressed by the CGF
energy density and pressure
\eqg
T^{\CGF}_{\mu\nu} = T^{\gauge}_{\mu\nu} +  T^{\phi}_{\mu\nu}
\\[2ex]
T^{\CGF}_{00} = \rho_{\CGF} (-g_{00})
\qquad
T^{\CGF}_{ij} = p_{\CGF} g_{ij}
\qquad
T^{\CGF}_{i0} = 0
\\[2ex]
\label{eq:densitypressurebefore}
\begin{alignedat}{2}
\frac1{\hbar}\rho_{\CGF} &= \frac{3 }{\gtwo ^{2}}\,\frac{\EhatCGF}{R^{4}}
+\frac{ \lambdaH^{2}\vH ^{4}}{8}
&&= \frac{3 }{8 \gtwo ^{2}}\,\frac{1}{a^{4}} 
+\frac{ 1}{8\lambdaH^{2}}
\,\frac{ 1}{ t_{\Higgs}^{4}}
\\[1ex]
\frac1{\hbar}p_{\CGF} &=\frac{1 }{\gtwo ^{2}}\,\frac{\EhatCGF}{R^{4}}
-\frac{ \lambdaH^{2}\vH ^{4}}{8}
&&= \frac{1 }{8 \gtwo ^{2}}\,\frac{1}{a^{4}} 
-\frac{ 1}{8\lambdaH^{2}}\,\frac{ 1}{ t_{\Higgs}^{4}}
\end{alignedat}
\eng
The first term in each expression is due the 
gauge field which is a perfect fluid with $w_{\gauge} = 1/3$.
The second term is due to the Higgs field vacuum energy,
a perfect fluid with $w_{\phi} = -1$.
The equation of state for $a\le a_{\EW}$ is
\eq
\label{eq:eqnofstatehigh1}
p_{\CGF} = \frac13 \rho_{\CGF} - \frac{ 1}{6\lambdaH^{2}}\,\frac{ 
\hbar}{ t_{\Higgs}^{4}}
\en

\subsection{Realizing the electroweak transition}
\label{sect:realizeEWtransition}
We have to show that the expanding universe driven by the
CGF and the Higgs field $\phi=0$ actually passes through $a=a_{\EW}$.
The Friedmann equation is
\eqg
H^{2}
+\frac1{R^{2}}
=
\frac13 \kappa \rho_{\CGF}
\qquad
H= \frac1{R} \frac{dR}{dt} = \frac1{R^{2}} \frac{dR}{d\that}
\eng
In terms of $z = \that/\epsilonb$, $a = \epsilonb R$, and 
$t_{\grav}^{2} = \hbar \kappa$
and using equation (\ref{eq:densitypressurebefore})
for the density,
\eqg
\label{eq:Friedmannequation}
H^{2}
+\frac{\epsilonb^{2}}{a^{2}}
=
\frac{1 }{8\gtwo ^{2}}
\,\frac{t_{\grav}^{2}}{a^{4}}
+ 
 \frac{ 1}{24\lambdaH^{2}}
\,\frac{ t_{\grav}^{2}}{ t_{\Higgs}^{4}}
\qquad
H= \frac1{a^{2}} \frac{da}{dz}
\eng
or, equivalently,
\eq
\label{eq:Friedmannequation2}
a^{2} H^{2} =
\frac{ t_{\grav}^{2}}{ t_{\Higgs}^{2}}
\left(
\frac{1 }{8\gtwo ^{2}}
\,\frac{t_{\Higgs}^{2}}{a^{2}}
+ \frac{ 1}{24\lambdaH^{2}}
\,\frac{ a^{2}}{ t_{\Higgs}^{2}}
\right)
-\epsilonb^{2}
\en
There is a solution
as long as the right hand side is positive.
Equation (\ref{eq:Friedmannequation2}) at $a=a_{\EW}$ is
\eq
\label{eq:FriedmannequationEW}
a_{\EW}^{2} H_{\EW}^{2} =
\frac{ t_{\grav}^{2}}{ t_{\Higgs}^{2}}
\left(
\frac{1 }{8\gtwo ^{2}}
\,\frac{t_{\Higgs}^{2}}{a_{\EW}^{2}}
+ \frac{ 1}{24\lambdaH^{2}}
\,\frac{ a_{\EW}^{2}}{ t_{\Higgs}^{2}}
\right)
-\epsilonb^{2}
=2.40\ntimes 10^{-33}-\epsilonb^{2}
\en
so the expansion passes through $a_{\EW}$ if
$\epsilonb^{2} < 2.40\ntimes 10^{-33}$.
The expansion passes through all values of $a$ if
the minimum of $a^{2}H^{2}$
in (\ref{eq:Friedmannequation2}) is positive,
\eq
\epsilonb^{2} < \frac{ t_{\grav}^{2}}{ t_{\Higgs}^{2}}
\left(
\frac{1 }{8\gtwo ^{2}}
\,\frac{ g}{ \sqrt3 \lambdaH}
+ \frac{ 1}{24\lambdaH^{2}}
\,\frac{ \sqrt3 \lambdaH}{ g}
\right)
=\frac{ t_{\grav}^{2}}{ t_{\Higgs}^{2}}
\frac{ 1}{4\sqrt3 g \lambdaH }
=1.15\ntimes 10^{-33}
\en
So we assume $\epsilon^{2} <  10^{-33}$ which is
$\EhatCGF > 10^{67}$
in order that the expansion passes through $a_{\EW}$
and  in order that the CGF oscillates for some considerable
period before $a_{\EW}$.

\subsection{Adiabatic condition for $a\le a_{\EW}$}
\label{sect:Adiabaticcond2}

In order to justify averaging over the CGF oscillation,
we need to show that 
the expansion time $1/H$ is much longer than the
oscillation period $4Ka$ for $a\le a_{\EW}$, i.e.
\eq
\label{eq:adiabaticcondition}
4K a H \ll 1
\en
For $a$ on the order of $a_{\EW}$,
equation (\ref{eq:FriedmannequationEW}) gives the ratio of time 
scales $4KaH$ to be on the order of $10^{-16}$
so the adiabatic condition (\ref{eq:adiabaticcondition}) is well 
satisfied.
For $a$ significantly less than $a_{\EW}$, the term in (\ref{eq:Friedmannequation2}) 
proportional to $1/a^{2}$ dominates, so the adiabatic condition
is satisfied as long as
\eq
(4K)^{2}a^{2}H^{2} \sim
(4K)^{2}
\frac{ t_{\grav}^{2}}{ t_{\Higgs}^{2}}
\left(\frac{1 }{8\gtwo ^{2}}
\,\frac{t_{\Higgs}^{2}}{a^{2}}\right)
\ll 1
\en
which is $a\, {\gg}\, t_{\grav}$
so there is plenty of time for expansion up to $a_{\EW}$
with the adiabatic condition well satisfied.

\subsection{Temperature from the CGF}

The cosmological gauge field $b(z)=k\cn(z,k)$ is an elliptic function 
of  the time $z$,
periodic in real time with period $4K$ and also periodic in imaginary time
with period $4K'\, i$.
The imaginary period in co-moving time defines an inverse temperature
$T_{\CGF}$.
\eq
\frac{\hbar}{\kB T_{\CGF}} =  4K' a 
\en
This is to be the origin of the cosmological temperature.
The CGF acts as thermal bath for the fluctuations
of the Standard Model fields.
The correlation functions of the fluctuations
are periodic in imaginary time with the period $ 4K' a \, i$.
This defines a specific semi-classical quantum state of the Standard Model
plus General Relativity
whose time evolution
describes the universe from $a_{\EW}$ onwards.

In \cite{Friedan2022:Stability} the initial thermal state of the 
$\SU(2)$ gauge field fluctuations is constructed
and shown to be thermodynamical stable,
a first step towards
showing that
the proposed $\Spin(4)$-symmetric initial condition is 
thermodynamically stable and thus physically 
natural.

\section{Mathematical literature}

The $\SO(4)$-invariant classical solution of pure $\SU(2)$ gauge theory on $S^{3}$
(without the Higgs field)
was first discovered as a mathematical object by
Alfaro, Fubini, and Furlan \cite{deAlfaro:1976qet},
Cervero, Jacobs, and Nohl \cite{Cervero:1977gm},
and L\"uscher \cite{Luscher:1977cw}.
Ivanova, Lechtenfeld, and Popov 
\cite{Ivanova:2017xjp,Ivanova:2017wun} examined
properties of the $\SO(4)$-invariant gauge theory solution on 
de Sitter and anti-de Sitter space.
Lechtenfeld described the back reaction
in General Relativity \cite{Lechtenfeld:2020wfl}.
The symmetry group was $\SO(4)$ in all of these works because 
$\Spin(4)$ symmetry can be posited only when there are
$\SU(2)$ doublet fields such as the Higgs field.
None of these mathematical works
made or suggested any connection between the mathematical solutions and 
physical cosmology or the electroweak transition
or thermal physics.
I thank O.~Lechtenfeld for bringing these mathematical works to my 
attention after the release of \cite{Friedan:2020poe}.

\section{Adiabatic time evolution of the CGF}

\subsection{Oscillating CGF at $a\ge a_{\EW}$}

As the scale $a(z)$ approaches $a_{\EW}$ the fluctuations of the
Higgs field around $\phi=0$ grow large.  After $a_{\EW}$ the Higgs 
field $\phi$ fluctuates in and around the minima of the effective potential 
(\ref{eq:VHiggseff}) which is
\eq
\label{eq:phi2}
\phi^{\dagger}\phi=
(\phi^{\dagger}\phi)_{0}
=
\frac{\vH ^{2}}2 
-
\frac3{4\lambdaH^{2}}  \frac{\expval{b^{2}} }{a^{2}}
\en
As $a(z)$ increases slowly, the potential well deepens and the 
$\phi$ fluctuations concentrate at $\phi^{\dagger}\phi=
(\phi^{\dagger}\phi)_{0}$.
The $\phi$ fluctuations are $\Spin(4)$-symmetric so the 
effective action remains $\Spin(4)$-symmetric.
The CGF continues to evolve as a $\Spin(4)$-symmetric $\SU(2)$ gauge field
$b(z) \gamma_{i}$ but now with effective action
\eqg
\label{eq:Seffgauge}
\frac1{\hbar} S^{\eff}_{\gauge} = \frac{\Vol(S^{3})}{\epsilonb^{3}}\frac{3  }{g^{2}}
\int
\bigg[-\frac12 \bigg(\frac{d b}{d z}\bigg)^{2}
+ \frac1{4} g^{2} a^{2} (\phi^{\dagger}\phi)_{0}b^{2} 
+ \frac12 b^{4}
\bigg]
dz
\eng
comprising the gauge action (\ref{eq:gaugeaction})
plus the $b$-dependent term of the $\phi$ action (\ref{eq:Higgsaction}).
After $a_{\EW}$ the CGF $b(z)$ oscillates in a quartic 
potential
whose quadratic term is changing very slowly compared to the 
oscillation.
The dimensionless conserved energy is
\eqg
\label{eq:Ebandmu}
\ECGF =  \epsilon^{4} \EhatCGF
= \frac12 \bigg(\frac{d b}{d z}\bigg)^{2}
+\frac{1}{2}\mu^{2}  b^{2}
+\frac12 b^{4}
\qquad
\mu^{2} = \frac12 \gtwo ^{2}a^{2}(\phi^{\dagger}\phi)_{0}
\eng
Again by equation~(\ref{eq:kcneqn})
the oscillating solution for fixed $\mu$ and $\ECGF$ is
\eq
b(z) = \frac{ k \cn(u,k) }\alpha\qquad  dz = \alpha du
\en
with parameters $k^{2}$, $\alpha$ depending on $\mu$, $\ECGF$ by
\eq
\label{eq:muEbofk}
\mu^{2}=\frac{1-2k^{2}}{\alpha^{2}}
\qquad
\ECGF  =\frac{ k^{2}(1-k^{2})}{2\alpha^{4}}
\en
$\mu^{2}\ge 0$ so $k^{2} \le 1/2$.
Equations (\ref{eq:phi2}) and (\ref{eq:Ebandmu}) determine $\mu$ in terms of
$\expval{b^{2}}$ which is the average over a period of 
oscillation,
\eq
\expval{b^{2}} = \frac1{4K} \int_{0}^{4K} b^{2} \,du
= \frac1{4K} \int_{0}^{4K} \frac1{\alpha^{2}}  k^{2} 
\cn^{2}(u,k) \; du
=\frac1{\alpha^{2}} \left( k^{2}-1 + \frac{E}{K} \right)
\en
evaluated using equation (\ref{eq:cnintegrals}).
$E=E(k)$ is the complete elliptic integral of the second kind.
We are now re-using the variables $k$ and $K=K(k)$.
From now on we will write
$k_{\EW}$, $K_{\EW}$, and $E_{\EW}$ for the values at $a\le a_{\EW}$,
the constants
\eq
k^{2}_{\EW} = \frac12
\qquad
K_{\EW} = K(k_{\EW})
\qquad
E_{\EW} = E(k_{\EW}) = \frac{K_{\EW}}{2} + \frac{\pi}{4 K_{\EW}}
\en
After $a_{\EW}$
the parameters $\mu$, $\ECGF$ and
$k$, $\alpha$ evolve slowly away from their values at $a_{\EW}$
\eq
\mu_{\EW}= 0
\qquad
E_{\CGF,\EW} =  \frac18
\qquad
k_{\EW}^{2} = \frac12
\qquad
\alpha_{\EW} = 1
\en
The equations derived so far,
\eqg
\label{eq:collected}
(\phi^{\dagger}\phi)_{0}=\frac{\vH ^{2}}2 
-\frac3{4\lambdaH^{2}}  \frac{\expval{b^{2}} }{a^{2}}
\qquad\quad
\mu^{2} = \frac12 \gtwo ^{2}a^{2}(\phi^{\dagger}\phi)_{0}
\\[2ex]
\mu^{2} =\frac{1-2k^{2}}{\alpha^{2}}
\qquad
\ECGF  =\frac{k^{2}(1-k^{2})}{2 \alpha^{4}}
\qquad
\expval{b^{2}} =\frac1{\alpha^{2}} \left( k^{2}-1 + \frac{E}{K} \right)
\eng
are one equation short of
determining the time evolution completely.

\subsection{Adiabatic invariant}

The adiabatic invariant
\eq
\label{eq:I}
I = \frac1{2\pi} \oint p \,dq
\en
is a constant of motion
in the adiabatic time evolution of an oscillating classical degree of 
freedom $q$ with conjugate momentum $p$.
The integral is over one oscillation.
Here, from the action (\ref{eq:Seffgauge}),
\eqg
q = b 
\qquad
\frac{1}{\hbar} p= \frac{\Vol(S^{3})}{\epsilonb^{3}}\frac{3  }{g^{2}}
\frac{d b}{d z}
\eng
so, using (\ref{eq:cnintegrals}),
\eqa
\label{eq:adiabaticinvariant}
\frac{1}{\hbar}I &=\frac1{2\pi} \frac{\Vol(S^{3})}{\epsilonb^{3}}\frac{3  }{g^{2}}
\int_{0}^{4K} \frac{d b}{d z} \frac{d b}{d u} \;du
=
\frac{\Vol(S^{3})}{2\pi\epsilonb^{3}}\frac{3  }{g^{2}}
\frac1{\alpha^{3}}
\int_{0}^{4K} k^{2} \cn'(u,k)^{2} \;du
\\[2ex]
&=\frac{\Vol(S^{3})}{2\pi\epsilonb^{3}}\frac{3  }{g^{2}}
\frac1{\alpha^{3}}
\frac43 \left[(1-k^{2})K + (2k^{2}-1)E\right]
\ena
Dropping the constant factors, we use
\eqa
\tilde I = \frac1{\alpha^{3}}\left[(1-k^{2})K + (2k^{2}-1)E\right]
\ena
as the adiabatic constant of motion.
At  $a=a_{\EW}$, $k^{2}_{\EW}=1/2$ and $\alpha =1$ so
\eq
\label{eq:tildeI}
\tilde I = \frac12 K_{\EW}
\en
so the adiabatic equation is
\eq
\label{eq:adiabaticequation}
\frac1{\alpha^{3}}\left[(1-k^{2})K + (2k^{2}-1)E\right] =
\frac12 K_{\EW}
\en
Now there are enough equations to 
parametrize the time evolution by $k^{2}$.

\subsection{$a$ as a function of the elliptic parameter $k^{2}$}

\begin{figure}
\begin{center}
\captionsetup{justification=centering,margin=0.03\linewidth}
\includegraphics[scale=0.8]{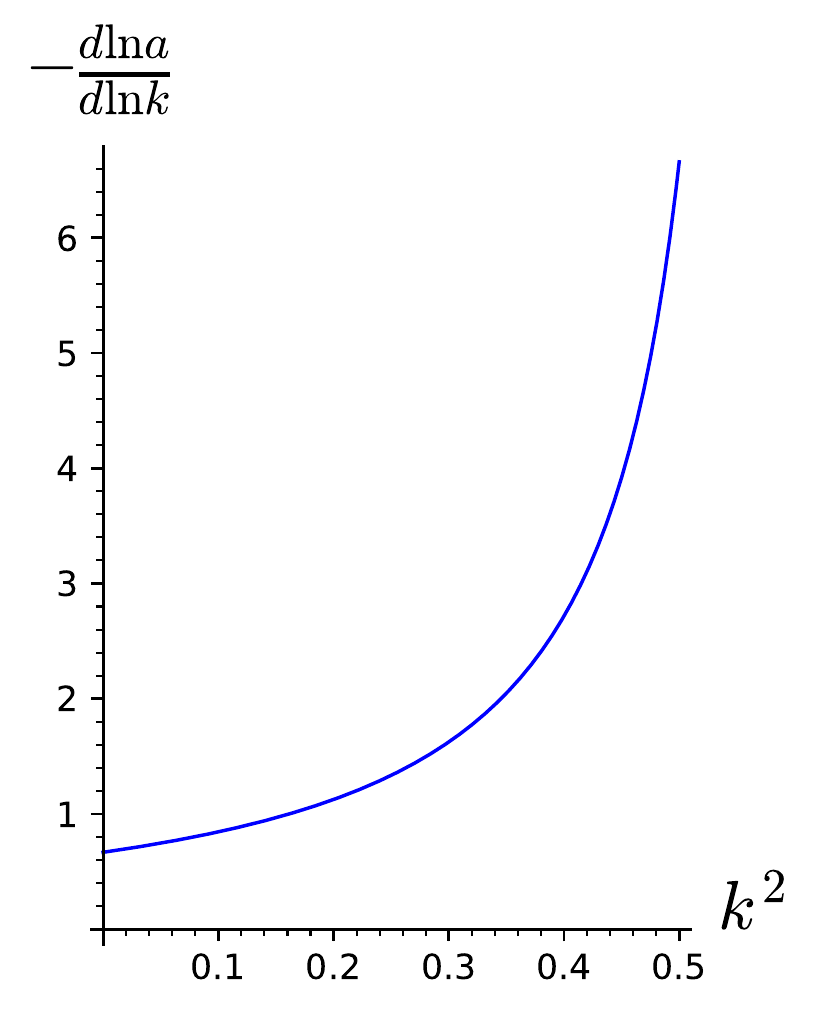}
\hfill
\includegraphics[scale=0.8]{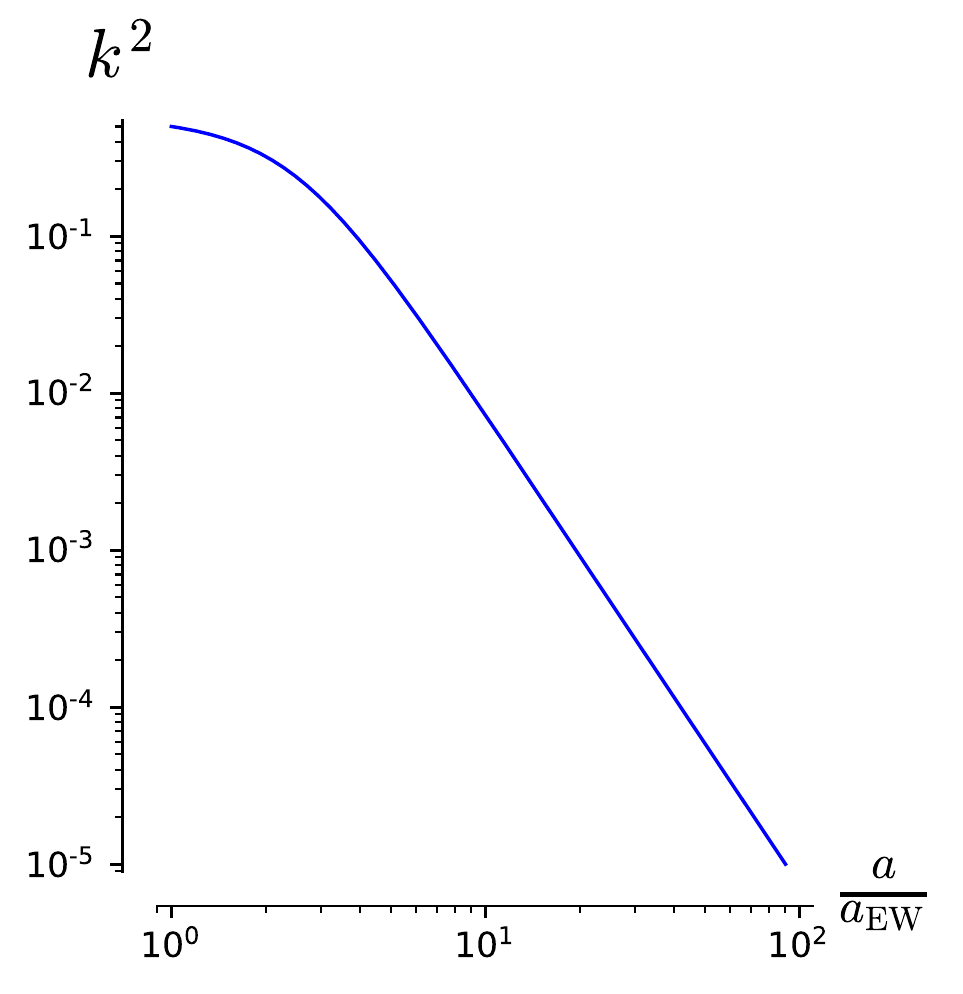}
\caption{The left graph shows that $a$
increases monotonically as $k^{2}$
decreases from $1/2$ to $0$.
The right graph shows that
the $k^{2}\rightarrow 0$  regime
is $a \ge 10^{2} a_{\EW}$.
\label{fig:monotonic}
}
\end{center}
\end{figure}

The adiabatic equation (\ref{eq:adiabaticequation}) gives $\alpha$ as a 
function of $k^{2}$.
Then (\ref{eq:collected}) gives $\mu^{2}$, $\ECGF$, and 
$\expval{b^{2}}$ as functions of $k^{2}$,
and then $a$ as a function of $k^{2}$,
\eq
\label{eq:aofk}
a =t_{\Higgs} \hat a 
\qquad
\hat a^{2}
= \frac3{2}  \expval{b^{2}}  + 
\frac{4\lambdaH^{2}}{\gtwo ^{2}}
\mu^{2}
\en
Numerical calculations graphed in
Figure~\ref{fig:monotonic} show that $a$ increases monotonically from $a_{\EW}$ to $\infty$ as 
$k^{2}$ decreases from $1/2$ to $0$.
So the parametrization by $k^{2}$
implicitly gives the time evolution  as a function of the scale $a$.
The $k^{2}\rightarrow 0$ regime is reached after about two ten-folds of 
expansion from $a_{\EW}$.

\subsection{CGF density and pressure after $a_{\EW}$}

For $a\ge a_{\EW}$,
the energy-momentum tensors (\ref{eq:Tmunu})
for the $\Spin(4)$-symmetric state in the classical and adiabatic 
approximation give
\eq
\rho_{\CGF} =  \frac\hbar{a^{4}}\left(\frac{3 E_{\CGF} }{\gtwo ^{2}}
+   \frac{9\expval{b^{2}}^{2} }{32\lambdaH^{2}} 
\right)
\qquad
 p_{\CGF} =  \frac\hbar{a^{4}}  \left(\frac{E_{\CGF}-\mu^{2} \expval{b^{2}} }{\gtwo ^{2}} 
- \frac{9\expval{b^{2}}^{2} }{32\lambdaH^{2} } \right)
\en
The dimensionless density and pressure are
\eq
\begin{alignedat}{2}
\hat\rho_{\CGF}  &= \frac{t_{\Higgs}^{4}\rho_{\CGF}}\hbar 
&&=  \frac{1}{\hat a^{4}}
\left(
\frac{3\ECGF}{\gtwo ^{2}} + \frac{9\expval{b^{2}}^{2}}{32\lambda^{2}}
\right )
\\[2ex]
\hat p_{\CGF}  &= \frac{t_{\Higgs}^{4} p_{\CGF}}\hbar
&&=  \frac1{\hat a^{4}}  \left(\frac{E_{\CGF}-\mu^{2} \expval{b^{2}} }{\gtwo ^{2}} 
- \frac{9\expval{b^{2}}^{2} }{32\lambdaH^{2} } \right)
\end{alignedat}
\en
These two equations
parametrize the density and pressure by $k^{2}$.
The equation of state relating $p$ to $\rho$ is given implicitly by the two equations.

\subsection{Parametrize the time evolution by $k^{2}$}

Table~\ref{table:parambyk} summarizes the 
parametrization by $k^{2}$.
The leading terms in the $k^{2}\rightarrow 0$ regime
are calculated in \cite{Friedan2022:AtheorySuppMat}
using the Taylor series expansions (\ref{eq:EKtaylorseries})
of $K$ and $E$.

\begin{table}
$$
{ \everymath={\displaystyle}
\def\arraystretch{2.5}
\label{eq:parametrizationbyk}
\begin{array}{|@{\quad}r@{\,}l@{\quad}|@{\quad}c@{\quad}l@{\;\;}l@{\quad}|}
\multicolumn{2}{c}{}& k_{\EW}^{2}=\frac12 
&\multicolumn{1}{l}{ k^{2}\rightarrow 0 }\\[1ex]
\hline
 \alpha^{3}  
&  {}\, = \frac{2(1-k^{2})K + 2 (2k^{2}-1) E  }{K_{\EW}}
&1&  \rightarrow\;\; \frac{3\pi k^{2}}{2K_{\EW}}  &   
\\[1.5ex]
\hline
\alpha^{2} \expval{b^{2}} 
&  {}\, =   k^{2}-1 + \frac{E}{K} 
&  \frac{\pi}{4K_{\EW}^{2}} &  \rightarrow\;\; \frac{k^{2}}{2}   
&   
\\[1.5ex]
\hline
\alpha^{2} \mu^{2} 
&  \,=1-2k^{2}
&  0  &  \rightarrow\;\; 1
&  
\\[1.5ex]
\hline
\alpha^{4} \ECGF 
&  \, =\frac{k^{2}(1-k^{2})}{2}  
&  \frac18  &  \rightarrow\;\;  \frac{k^{2}}{2} 
&    
\\[1.5ex]
\hline
\alpha^{2} \hat a^{2} 
&  \,= \frac{3 \alpha^{2}\expval{b^{2}}}{2} + \frac{4\lambdaH^{2}\alpha^{2}\mu^{2} }{\gtwo ^{2}}
&  \frac{3\pi}{8 K_{\EW}^{2}} &  \rightarrow\;\;  \frac{4\lambda^{2}}{\gtwo^{2}} 
&    
\\[1.5ex]
\hline
\hat\rho_{\CGF} 
&  \,=  \frac1{\hat a^{4}} 
\left(
\frac{3\ECGF}{\gtwo ^{2}} + \frac{9\expval{b^{2}}^{2}}{32\lambda^{2}}
\right )
&  \frac{8 K_{\EW}^{4}}{3 \pi^{2}g^{2}} +\frac{1}{8\lambda^{2}}  &  \displaystyle\rightarrow\;\;  \frac{3g^{2}k^{2}}{32\lambda^{4}} 
&    
\\[1.5ex]
\hline
\hat p_{\CGF} 
&  \,=  \frac1{\hat a^{4}}  \left(\frac{E_{\CGF}-\mu^{2} \expval{b^{2}} }{\gtwo ^{2}} 
- \frac{9\expval{b^{2}}^{2} }{32\lambdaH^{2} } \right)
& \frac{8 K_{\EW}^{4}}{9 \pi^{2}g^{2}} -\frac{1}{8\lambda^{2}}  &  \displaystyle\rightarrow\;\;  
\frac{9 g^{2}(8\lambda^{2}-g^{2})k^{4}}{2048\lambda^{6}}
&   
\\[1.5ex]
\hline
 \frac{2(\phi^{\dagger}\phi)_{0}}{\vH^{2}}   
&  \,=1-\frac{3\expval{b^{2}} } {2 \hat a^{2} }  
&  0  & 
\multicolumn{1}{l}{\!\!\! \rightarrow\;\;  1 -\frac{3 \gtwo^{2} 
k^{2}}{16\lambdaH^{2}}  \hfill}& 
\\[1ex]\hline
\end{array}
}
\def\arraystretch{1}
$$
\caption{
The parametrization by $k^{2}$.
\label{table:parambyk}
}
\end{table}


\section{CGF as the dark matter}

\subsection{$\Omega_{\CGF}+\Omega_{\Lambda}=1$ in the present}

The Friedmann equation is
\eq
H^{2} + \frac1{R^{2}} = \frac13 \kappa (\rho_{\ssm}+\rho_{\ssLambda})
\qquad
H = \frac1{R} \frac{dR}{dt}
\en
$\rho_{\ssm}$ is the matter density
and $\rho_{\ssLambda}$ is the dark energy density (assumed due to the 
cosmological constant).
In scaled coordinates,
\eqg
\label{eq:Friedmanneqn}
R = \frac{a}{\epsilon}
\qquad
dt = a dz
\\[2ex]
H^{2} + \frac{\epsilonb^{2}}{a^{2}} = \frac13 \kappa (\rho_{\ssm}+\rho_{\ssLambda})
\qquad
H = \frac1{a^{2}} \frac{da}{dz}
\qquad
\eng
We are ignoring fluctuations so the entire matter density is
\eq
\rho_{\ssm} = \rho_{\CGF}
\en
The Friedmann equation normalized by $H_{0}^{2}$ is
\eqg
\label{eq:Friedmanneqnnormalized}
\frac{H^{2}}{H_{0}^{2}}  = \Omega_{\CGF} + 
\Omega_{\ssLambda}+\Omega_{\curvature} 
\\[1ex]
\rho_{\ssc} = \frac{3}{\kappa} H_{0}^{2}
\qquad
\Omega_{\CGF} = \frac{\rho_{\CGF}}{\rho_{\ssc}}
\qquad
\Omega_{\ssLambda} = \frac{\rho_{\ssLambda}}{\rho_{\ssc}}
\qquad
-\Omega_{\curvature} 
=\frac{1}{H_{0}^{2}}\frac{1}{R^{2}}=\frac{ t_{\Hubble}^{2}\epsilon^{2}}{a^{2}}
\eng
$\rho_{\ssc}$ is the critical density.  
The observed present curvature is close to zero,
$|\Omega_{\curvature}|<0.001$.
The observed dark energy density
is $\Omega_{\ssLambda}=0.685$.
We assume that the dark energy is due to the cosmological constant,
i.e. that $\Omega_{\ssLambda}$ is constant in time.

Using the $k\rightarrow 0$ asymptotic formulas,
\eq
\Omega_{\CGF}
= \frac{\kappa}{3H_{0}^{2}} \rho_{\CGF} 
= \frac{t_{\Hubble}^{2}t_{\grav}^{2}}{3 t_{\Higgs}^{4}} \hat\rho_{\CGF} 
= 
\frac{t_{\Hubble}^{2}t_{\grav}^{2}}{t_{\Higgs}^{4}}
\frac{g^{2}k^{2}}{32\lambda^{4}}
=
\frac{K_{\EW}}{6\pi\gtwo\lambda }
\frac{t_{\Hubble}^{2}t_{\grav}^{2}}{t_{\Higgs}a^{3}}
\en
The present time is identified by the condition $H=H_{0}$.
By the Friedmann equation (\ref{eq:Friedmanneqnnormalized}),
$H=H_{0}$ is  equivalent to
$\Omega_{\CGF} = 0.315$
assuming that $|\Omega_{\curvature}|<0.001$.
So the present values of $k^{2}$ and $a$ are
\eqa
 k_{0}^{2} 
&=  0.315 \, 
\frac{t_{\Higgs}^{4}}{t_{\Hubble}^{2}t_{\grav}^{2}}
\frac{32\lambda^{4}} {g^{2}}
= 7.89\ntimes 10^{-56}
\\[2ex]
a_{0}
&= 
\left( 
0.315 \, 
\frac{6\pi\gtwo\lambda }{K_{\EW}}
\frac{t_{\Higgs}}{t_{\Hubble}^{2}t_{\grav}^{2}}
\right)^{-\frac13}
\\[1ex]
&= 1.40\ntimes 10^{-8} \,\sunit
 = 2.66 \ntimes 10^{18} t_{\Higgs}
=  4.54 \ntimes 10^{18} a_{\EW}
\ena
$k_{0}^{2}$ is very small so using the asymptotic formulas is justified.
The present curvature is
\eq
-\Omega_{\curvature} 
= t_{\Hubble}^{2} \frac{\epsilon^{2}}{a_{0}^{2}}
= 1.07\ntimes 10^{51} \; \epsilon^{2} 
\en
so the present flatness condition, $|\Omega_{\curvature}| < 0.001$, is
equivalent to
$\epsilon  <  10^{-27}$
which is
$\EhatCGF >  10^{107}$.
\eq
\label{eq:EhatCGFbound}
\EhatCGF >  10^{107}
\quad\Longleftrightarrow\quad
|\Omega_{\curvature}| < 0.001
\en
Then
\eq
\Omega_{\CGF}+\Omega_{\Lambda} =1
\en
The CGF is the dark matter.

\subsection{$w_{\CGF}\approx 0$ after $10^{2}a_{\EW}$}

\begin{figure}[H]
\begin{center}
\captionsetup{justification=centering,margin=0.19\linewidth,skip=0pt}
\includegraphics[scale=0.8]{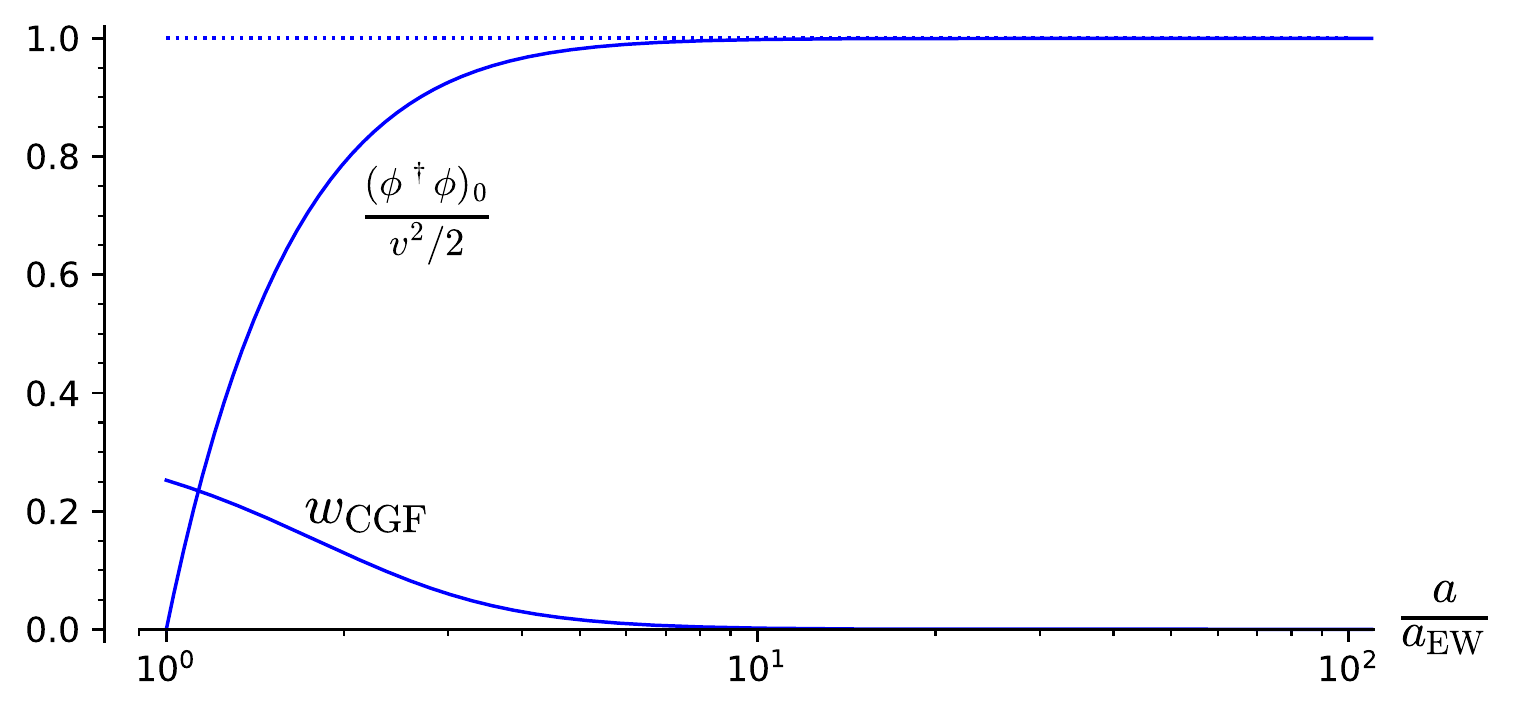}
\caption{
}
\label{fig:wandphisq}
\end{center}
\end{figure}

The equation of state parameter is
\eq
w_{\CGF} = \frac{p_{\CGF}}{\rho_{\CGF}}
\en
The formulas of Table~\ref{table:parambyk} give,
in the $k\rightarrow 0$ regime,
\eq
w_{\CGF} = \left(\frac38 - \frac{3g^{2}}{64\lambda^{2}}\right) k^{2}
\en
so the CGF evolves as a non-relativistic fluid ($w=0$)
after the first one or 
two ten-folds of expansion from $a_{\EW}$.
Figure~\ref{fig:wandphisq}
shows $w_{\CGF}$ approaching 0 as the electroweak 
transition approaches completion,
i.e. as $(\phi^{\dagger}\phi)_{0}$
approaches the vacuum expectation value $\vH ^{2}/2$.

\vskip-2ex
\noindent
In the $k\rightarrow 0$ regime
the formulas of Table~\ref{table:parambyk} give
\eq
\rho_{\CGF}
=
 \frac{K_{\EW}}{2\pi \gtwo\lambda}\,
\frac{\hbar }{t_{\Higgs}}
\frac1{a^{3}}
=  \frac{0.890\,m_{\Higgs}}{a^{3}}
\en
showing the $1/a^{3}$ behavior of a $w=0$ fluid,
which follows from the conservation of energy-momentum
\eqg
\dot \rho+ 3 \frac{\dot a}{a} (\rho + p) = 0
\qquad
\frac{d\ln \rho}{d\ln a} = -3(1+w)
\eng

\subsection{The CGF in the present}
Let $x_{\phys}^{\mu}$ be 
the dimensionful physical coordinates 
\eq
x_{\phys}^{0} = t \qquad x_{\phys}^{i}= a x^{i}
\qquad
ds^{2} = dt^{2}- \delta_{ij} dx_{\phys}^{i}dx_{\phys}^{j}
\en
The CGF is
\eqg
B_{\CGF} =
B_{\CGF,i}^{\phys} dx_{\phys}^{i}
\eng
The dimensionful gauge field $B_{\CGF,i}^{\phys}$ is given by
\eq
B_{\CGF,i}^{\phys} dx_{\phys}^{i}
=
b(z) \gamma_{i}d x^{i}
\en
which is
\eq
B_{\CGF,i}^{\phys}=\frac{b(z)}{a} \gamma_{i}
=\frac{k\cn(u,k)}{\alpha a} \gamma_{i} 
\qquad
du = \frac1{\alpha} dz = \frac1{\alpha a}dt
\en
In the $k\rightarrow 0$ regime ($a > 10^{2} a_{\EW}$), 
\eq
\alpha a = \frac{2\lambdaH t_{\Higgs}}g = 
\frac{2\lambdaH \hbar}{g m_{\Higgs}}
 =  \frac{\hbar}{m_{W}}
=
\frac1{\omega_{W}}
\qquad
\cn(u,k) = \cos(u) = \cos(\omega_{W}t)
\en
so the dimensionful CGF is
\eq
B_{\CGF,i}^{\phys}= k \omega_{W} \cos(\omega_{W}t)\gamma_{i}
\qquad
\omega_{W}= \frac{g}{2\lambdaH t_{\Higgs}} = \frac{m_{W}}{\hbar} = 1.22\ntimes 10^{26}\,\sunit^{-1}
\en
The present CGF oscillates harmonically
at the bottom of the massive gauge field potential.
The oscillation period is
\eq
4 K \alpha a
= \frac{2\pi}{\omega_{W}} = 5.15\ntimes 10^{-26}\,\sunit
\en
The present value of $k$ is $k_{0} = 2.81 \ntimes 10^{-28}$ so, 
presently,
\eq
B_{\CGF,i}^{\phys}= 2.26\ntimes 10^{-26}\,\frac{\GeV}{\hbar} \cos(\omega_{W}t)\gamma_{i}
\en
Again, this is a leading order calculation, without fluctuations.
Higher order effects presumably cause
the fluctuating CGF to collapse gravitationally into an 
ensemble of self-gravitating bodies such as
the dark matter stars described in \cite{Friedan2022:DMStars}.

\subsection{CGF equation of state}

\begin{figure}
\begin{center}
\captionsetup{justification=centering,margin=0.03\linewidth}
\includegraphics[scale=0.8]{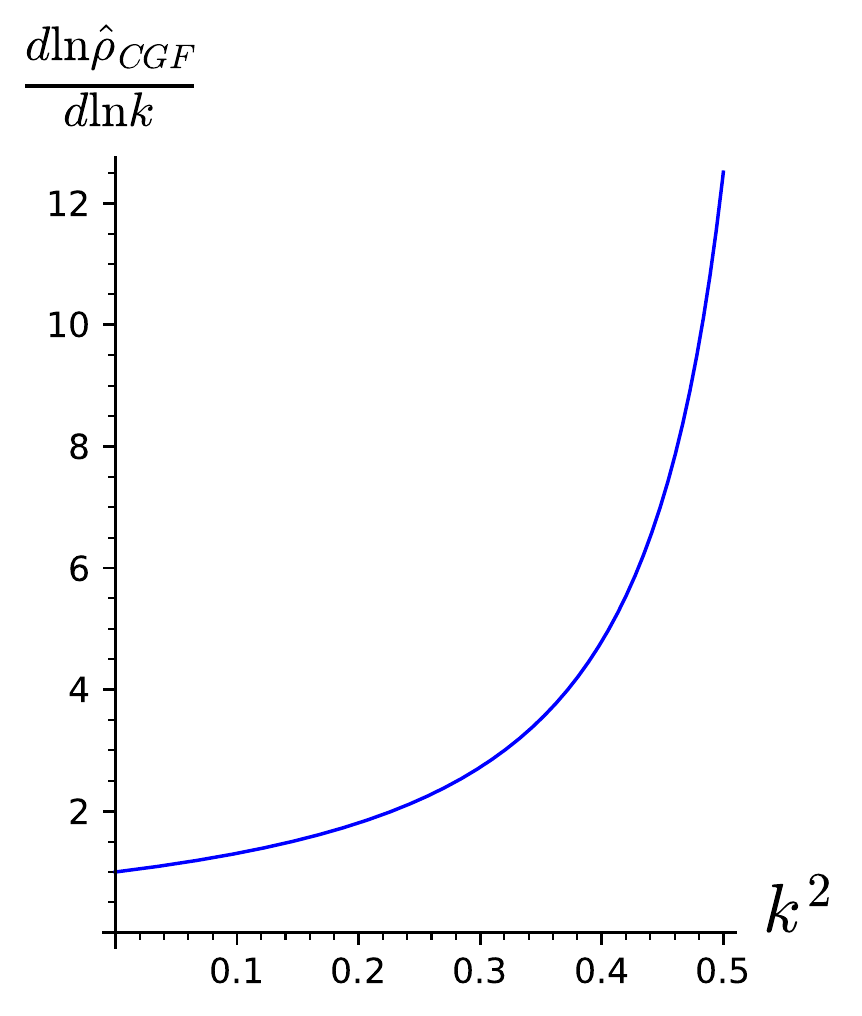}
\caption{$\hat\rho_{\CGF}$
increases monotonically with $k^{2}$.
\label{fig:rhomonotonic}
}
\end{center}
\end{figure}

The density scale of the CGF is
\eq
\rho_{b} = \frac{\hbar}{t_{\Higgs}^{4}}
= 5.68 \ntimes 10^{28} \, \frac{\kg}{\munit^{3}}
\en
The dimensionless CGF density and pressure are
\eq
\hat \rho_{\CGF} = \frac{\rho_{\CGF}}{\rho_{b}}
\qquad
\hat p_{\CGF} = \frac{p_{\CGF}}{\rho_{b}}
\en
The dimensionless density at $a_{\EW}$ from Table~\ref{table:parambyk} is
\eq
\hat \rho_{\EW} = \frac{8 K_{\EW}^{4}}{3 \pi^{2}g^{2}} +\frac{1}{8\lambda^{2}}
= 7.97
\en
Equation (\ref{eq:eqnofstatehigh1}) gives the equation of state for 
$\hat \rho\ge \hat \rho_{\EW}$
\eq
\hat\rho \ge \hat\rho_{\EW}
\qquad
\hat p 
= \frac13 \left( \hat \rho - c_{b}\hat\rho_{\EW}\right)
\qquad
c_{b} = \frac{ 1}{2\lambdaH^{2}\hat\rho_{\EW}} = 0.243
\en
For $\hat\rho \le \hat\rho_{\EW}$ the equation of state is given 
implicitly by the  analytic functions  of Table~\ref{table:parambyk}.
\eq
\hat\rho \le \hat\rho_{\EW}
\qquad
\hat \rho,\,\hat p =
\hat\rho_{\CGF}(k^{2}),\,\hat p_{\CGF}(k^{2})
\qquad
0 \le k^{2} \le \frac12
\en
The equation of state is well-defined because $\hat\rho_{\CGF}(k^{2})$
is monotonic in $k^{2}$ as shown in Figure~\ref{fig:rhomonotonic}.
The limit $k^{2}\rightarrow 0$ is 
$\hat\rho \rightarrow 0$.  From Table~\ref{table:parambyk}, the 
equation of state in the limit is
\eq
\rho \rightarrow 0
\qquad
p = \frac{c_{a}}{2}\rho^{2}
\qquad
c_{a} = \frac{ \lambda^{2}(8\lambda^{2}-g^{2})}{g^{2}}
= 0.992
\en

\subsection{Adiabatic condition for $a \ge a_{\EW} $}
\begin{figure}[H]
\begin{center}
\captionsetup{justification=centering,margin=0.2\linewidth}
\includegraphics[scale=0.8]{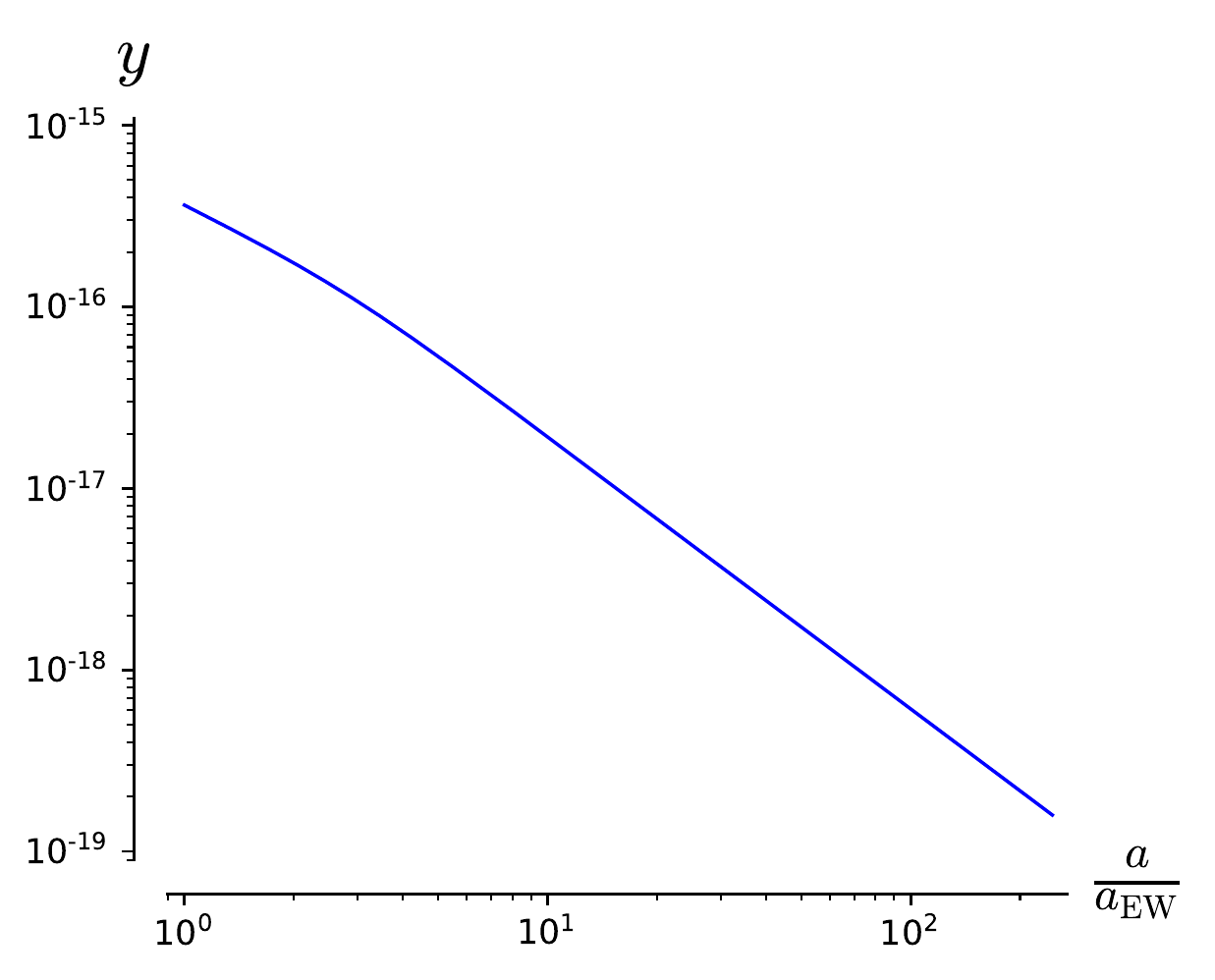}
\caption{The upper bound 
$y=4 K \alpha \sqrt{a^{2}H^{2}+\epsilon^{2}}$
on the ratio of time scales
$4 K \alpha a H$.
}
\label{fig:adiabaticbound}
\end{center}
\end{figure}

Finally, we need to verify that the adiabatic condition is satisfied for 
$a \ge a_{\EW} $, that the adiabatic approximation of the 
time evolution is justified.
The ratio of the oscillation period $4 K \alpha a$ to the expansion 
time $1/H$ is
$4 K \alpha a H $.
The Friedmann equation (\ref{eq:Friedmanneqn}) can be written
\eqg
\label{eq:Friedmann2}
a^{2}H^{2} + \epsilonb^{2} = \frac{t_{\grav}^{2}}{t_{\Higgs}^{2}} 
\frac13 \hat a^{2} \hat \rho_{\CGF}+ 
\frac{t_{\Higgs}^{2}}{t_{\Hubble}^{2}}
\Omega_{\Lambda}\hat a^{2}
\\[1ex]
\hat a = \frac{a}{t_{\Higgs}}
\qquad
\hat \rho_{\CGF} = \frac1\hbar 
t_{\Higgs}^{4}\rho_{\CGF}
\eng
The quantity $4 K \alpha \sqrt{a^{2}H^{2}+\epsilon^{2}}$
is an upper bound on the ratio of time scales $4 K \alpha a H$.
It is plotted in Figure~\ref{fig:adiabaticbound} for the first two 
ten-folds of expansion after $a_{\EW}$.
\noindent
At $a_{\EW}$ the bound is $3.64\ntimes 10^{-16}$.
The $\epsilon^{2}$ term is negligible, so this is the actual ratio of 
time scales at $a_{\EW}$.
The Friedmann equation (\ref{eq:Friedmann2}) in
the $k\rightarrow 0$ regime gives
\eqa
(4K\alpha a H)^{2} +(4K \alpha \epsilonb)^{2}
&= 
 \frac{t_{\grav}^{2}}{t_{\Higgs}^{2}} 
\frac{ \pi^{2}}{2 \lambdaH^{2}}k^{2}
+
\frac{t_{\Higgs}^{2}}{t_{\Hubble}^{2}}
 \frac{16\pi^{2}\lambdaH^{2}}{\gtwo ^{2}}\Omega_{\Lambda}
\\[1ex]
&=
5.04\ntimes 10^{-32} k^{2} 
+ 8.65\ntimes 10^{-87}
\ena
so the ratio of time scales $4 K \alpha a H $ 
decreases monotonically from $ 10^{-16}$ as $a$ increases 
from $a_{\EW}$.
The present ratio of time scales is
\eq
4 K_{0} \alpha_{0} a_{0} H_{0} = 2\pi \frac{2\lambdaH }{\gtwo }
\frac{t_{\Higgs} }{t_{\Hubble}}
= 1.12 \ntimes 10^{-43}
\en
The adiabatic condition is well satisfied from $a_{\EW}$ onward (and, 
as we saw earlier,
for some ten-folds before $a_{\EW}$).
The adiabatic approximation of the CGF time evolution is justified.

\section{Questions and comments}

\begin{center}
{
\begin{tabular}{r@{\:\:\:}l}
\ref{subsect:Test}& \hyperref[subsect:Test]{Testing the CGF cosmology} \\
\ref{subsect:Spin4}& \hyperref[subsect:Spin4]{Spin(4) and fluctuations} \\
\ref{subsect:Thermalization}& \hyperref[subsect:Thermalization]{Thermalization before $a_{\EW}$} \\
\ref{subsect:Decoupling}& \hyperref[subsect:Decoupling]{CGF temperature after  $a_{\EW}$} \\
\ref{subsect:CP}& \hyperref[subsect:CP]{CP violation} \\
\ref{subsect:Adiabatic}& \hyperref[subsect:Adiabatic]{Semi-classical approximation} \\
\ref{subsect:Neutrinos}& \hyperref[subsect:Neutrinos]{Neutrinos} \\
\end{tabular}
}
\end{center}

\subsection{Testing the CGF cosmology}
\label{subsect:Test}

Two approaches to checking the theory seem obvious.
\begin{enumerate}
\item
The initial thermal state of the fluctuations should be constructed
and its time evolution calculated to see if 
the right amount of ordinary matter results.
As a start, methods are developed in \cite{Friedan2022:Stability} to construct the initial 
thermal state of the fluctuations.

\item
If the CGF is indeed the dark matter, 
can it be detected?
What form does it take in the present?
What are its interactions with ordinary matter?
It seems reasonable to suppose that fluctuations in the CGF  
have collapsed gravitationally to stable 
self-gravitating structures.
As a first step, the Tolman-Oppenheimer-Volkoff stellar structure equations
for stars made of the CGF are solved numerically in \cite{Friedan2022:DMStars}.
It should be possible to model a purely dark matter universe
populated with galaxies of dark matter stars.
The actual universe would be a perturbation of the dark matter 
universe.
\end{enumerate}

If the theory survives testing then first principles 
cosmology will become possible.
All of the Standard Model cosmological epoch will be calculable from 
first principles.
Eventually, discrepancies between the theory and cosmological 
observation can become
clues to more fundamental physical principles.

\subsection{Spin(4) and fluctuations}
\label{subsect:Spin4}

The Spin(4) symmetry is a global symmetry of the universe
in the CGF cosmology.  
Presumably we live in a fluctuation where the global Spin(4) is not 
apparent.  We see the Poincar\'e symmetry of the Standard Model with
the Higgs field $\phi$ transforming as a scalar, not a spinor.

Before the electroweak transition,
as the universe expands towards $a_{\EW}$, the fluctuations of $\phi$
around $0$ grow large.
After $a_{\EW}$ the $\phi$ fluctuations concentrate again, 
now at the bottom of the effective Higgs potential, the set
$\phi^{\dagger}\phi= (\phi^{\dagger}\phi)_{0}$
which evolves slowly towards the vacuum expectation value
$v^{2}/2$.

Spin(4) continues to be a global symmetry of the universe.  The fluctuations of 
$\phi(x)$ at the bottom of the effective Higgs potential are 
Spin(4)-symmetric.  The crucial quantity is the  correlation 
length of the $\phi$ fluctuations after the electroweak transition.  This is the largest 
distance $\xi$ such that
\eq
\mathrm{dist}(x_{1},x_{2})< \xi
\quad
\implies
\quad
\expval{\phi^{\dagger}(x_{1})\,\phi(x_{2})} = \frac{v^{2}}{2}
\en
The Standard Model as we observe it
operates within regions of size $\xi$.

\subsection{Thermalization before $a_{\EW}$}
\label{subsect:Thermalization}

The imaginary time periodicity of the CGF defines a specific thermal state 
of the fluctuations of the Standard Model fields.
That state is thermodynamically stable against
fluctuations of the $\SU(2)$ gauge field \cite{Friedan2022:Stability}.
It remains to be checked that it is stable against all fluctuations.
A generic state presumably thermalizes to this state.
Is it possible to determine
how long the thermalization takes?
how long before $a_{\EW}$ the 
CGF must have been oscillating to ensure
that the fluctuations are in the thermal state at $a_{\EW}$?

\subsection{CGF temperature after $a_{\EW}$}
\label{subsect:Decoupling}

My first motivation for formulating the CGF cosmology
was to explain the origin of cosmological temperature
as the  imaginary time periodicity of the CGF \cite{Friedan:2020poe}.
I made a handwaving estimate of the redshift $a_{0}/a_{\EW}$
by assuming that the CGF temperature $T_{\CGF}$ at $a_{\EW}$ redshifted to the CMB 
temperature $T_{\CMB}$ in the present.
We can be  more definite
now that we know the time evolution of the CGF  after $a_{\EW}$.

The imaginary period in the time variable $u$ is $\Delta u = 4K'i$,
so the imaginary period in $z$ is $\Delta z = \alpha \Delta u = 
4K'\alpha i$,
so the imaginary period in co-moving time $t$ is
$\Delta t = a \Delta z = 4K'\alpha a i$.
So the inverse CGF temperature after $a_{\EW}$ is
\eq
\frac{\hbar}{\kB T_{\CGF}} = 4K'\alpha a 
\en
The present CGF temperature is
\eq
T_{\CGF,0} = \frac{\hbar}{\kB} \frac1{4K'_{0}\alpha_{0} a_{0}} = 
3.60 \ntimes 10^{12}\,\Kunit
\qquad
\kB T_{\CGF,0} = 310\text{\,MeV}
\en
so the ordinary matter must have decoupled from the CGF some time in 
the past.

\begin{figure}[H]
\begin{center}
\captionsetup{justification=centering,skip=0pt}
\includegraphics[scale=0.75]{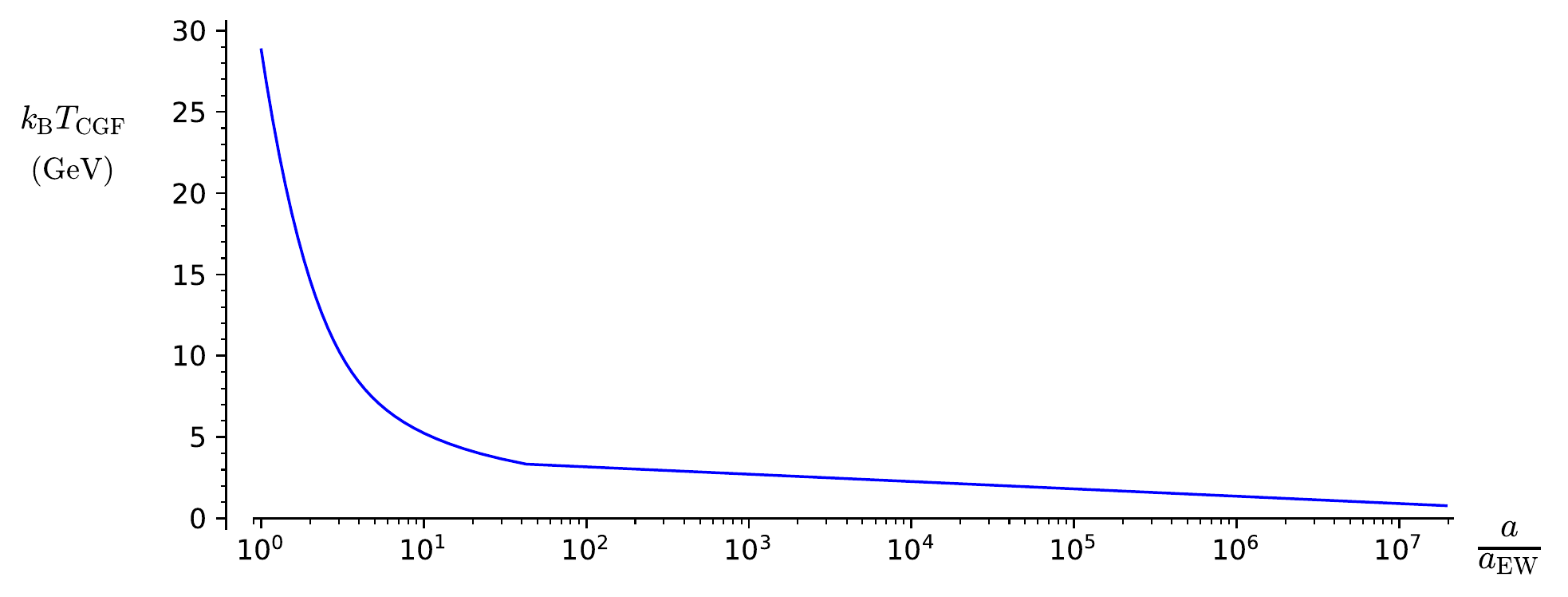}
\caption{}
\label{fig:TCGF}
\end{center}
\end{figure}

Figure~\ref{fig:TCGF} shows the CGF temperature after $a_{\EW}$.
\vskip-2ex
\noindent
In the $k\rightarrow 0$ regime, the asymptotic formulas and
equation~(\ref{eq:EKtaylorseries}) give
\eq
T_{\CGF} =
\frac{\hbar}{\kB} \frac1{4\ln(4/k)} \frac{g}{2\lambda t_{\Higgs}}
= \frac{2.33\ntimes 10^{14}\,\Kunit}{\ln(4/k)}
\qquad
\kB T_{\CGF}
= \frac{20.1\text{\,GeV}}{\ln(4/k)}
\en
Redshifting
the asymptotic CGF temperature $T_{\CGF}$ 
to the present gives
\eq
\frac{a}{a_{0}} T_{\CGF} = \frac{\hbar}{\kB} \frac1{4\ln(4/k)} \frac{g}{2\lambda t_{\Higgs}}\frac{a}{a_{0}}
= 
8.77\ntimes 10^{-5}\,\Kunit 
\;\frac{\hat a}{\ln(4/k)}
\en
The redshifted CGF temperature  will equal the CMB temperature,
\eq
\frac{a}{a_{0}} T_{\CGF} = T_{\CMB} = 2.7255\,\Kunit
\en
when $\hat a \approx 10^{6}$ which is
well within the asymptotic $k\rightarrow 0$ regime.
Numerical solution of 
\eqg
\frac{a}{a_{0}} T_{\CGF} 
= \frac{\hbar}{\kB} 
\frac1{4K'}\frac1{\alpha a_{0}}
=\frac{1.36\ntimes 10^{-4}\,\Kunit}{K' \alpha}
=2.7255\,\Kunit
\eng
finds \cite{Friedan2022:AtheorySuppMat}
\eqg
k^{2} = 5.12\ntimes 10^{-18}
\qquad
\frac{a}{a_{\EW}} = 1.13 \ntimes 10^{6}
\qquad
\kB T_{\CGF} = 0.944\,\GeV
\eng
which is indeed within the asymptotic regime.
The CGF has to decouple from the ordinary matter
at $a = 10^{6} a_{\EW}$ when $\kB T_{\CGF} = 1\,\text{GeV}$.

This is a very simplistic calculation.  It ignores fluctuations and 
supposes
that all of the ordinary matter decouples at the same time
and that the temperature simply redshifts from that time on.

\subsection{CP violation}
\label{subsect:CP}

If the CGF cosmology is to succeed as a complete theory of the 
Standard Model epoch,
it will have to explain the baryon-antibaryon asymmetry.
Everything after $a=a_{\EW}$ is completely
determined by the $\Spin(4)$-symmetric initial condition
leading up to the electroweak transition.
What comes before the Standard Model epoch is immaterial.

My thought in \cite{Friedan:2020poe} was that the discrete \CP 
symmetry takes 
$\hat 
b \rightarrow -\hat b$ (because the Dirac matrices $\gamma_{i}$ 
change sign when the orientation of space is reversed).  During the transition $\hat b$ 
settles in one of the minima $\hat b=\pm 1$ of the double well potential
(\ref{eq:baction}), breaking \CP as it settles.
But the two minima are indistinguishable
in local coordinates, i.e.~$b=\pm \epsilon$.
So any \CP violating effects have to be global.

The Higgs field fluctuations play a role in this.
The symmetry group $\Spin(4)$ is $\SU(2)\times\SU(2)$.
The Higgs field $\phi(x)$ decomposes into a sum of irreducible 
representations 
\eq
(1/2,0) \otimes \sum_{j} (j,j) = \sum_{j} (j+1/2,j) \oplus (j,j+1/2)
\en
\CP exchanges $(j+1/2,j) \leftrightarrow (j,j+1/2)$.
The orientation of the fluctuation in
the vector space $(j+1/2,j) \oplus (j,j+1/2)$
can break $\CP$.
Again, any effects will be global since $(j+1/2,j)$ and $(j,j+1/2)$
become indistinguishable when $j$ is large.

The expectation value of the baryon number density will be zero.
The question again is whether the baryon number density
is correlated over large enough regions and whether the magnitude of the two point function in such 
regions is big enough.

\subsection{Semi-classical approximation}
\label{subsect:Adiabatic}

A quantitative measure of the validity of the semi-classical approximation
is the adiabatic invariant $I$ of equation (\ref{eq:I})
An oscillating quantum system is semi-classical when $\frac1\hbar I 
\gg 1$.
For the CGF, equations (\ref{eq:adiabaticinvariant}) and 
(\ref{eq:tildeI}) and the bound (\ref{eq:EhatCGFbound}) on $\EhatCGF$ give
\eq
\frac1\hbar I =
\frac{\Vol(S^{3})}{2\pi\epsilonb^{3}}\frac{2  K_{\EW}  }{g^{2}}
> 10^{82}
\en
So  the CGF is semi-classical.

\subsection{Neutrinos}
\label{subsect:Neutrinos}

It is not entirely accurate
that the theory assumes nothing beyond the Standard Model.
This is true in the classical approximation.
When fluctuations are taken into account,
a mechanism will have to be added to the Standard Model
to produce neutrino masses and mixing.


\begin{appendix}
\section{Elliptic integrals and elliptic functions}
\label{app:elliptic}

The following formulas are from Gradsteyn and Ryzhik (G\&R) \cite{GR7th}
and the Digital Library of Mathematical Functions (DLMF) 
\cite{DLMF2022}:


\vskip2ex
\noindent
{\bf G\&R 8.11, DLMF 19.2(ii)}

\vskip1ex
The complete elliptic integrals of the first and second kinds are
\eqg
\label{eq:ellipticintegralsKE}
K(k) = \int_{0}^{\frac\pi2} \frac1{\sqrt{1-k^{2}\sin^{2}\alpha}} \; d\alpha
\qquad
E(k) = \int_{0}^{\frac\pi2} \sqrt{1-k^{2}\sin^{2}\alpha} \; d\alpha
\\[2ex]
k^{2}+k'{}^{2} =1
\qquad
K = K(k)
\quad
E = E(k)
\qquad
K' = K(k')
\quad
E' = E(k')
\eng
\noindent
$k$ is called the  {\it elliptic modulus}, $k'$ the {\it complementary 
modulus}.
$k^{2}$ is called the {\it elliptic parameter} (usually written $m$).
We can assume $0< k,k'<1$.

\vskip3ex
\noindent
{\bf G\&R 8.122, DLMF 19.7.1}
\eq
E K'  +E'  K  -K K'  = \frac\pi2
\en

\vskip2ex
\noindent
{\bf G\&R 8.129}
\eq
k^{2}=\frac12
\qquad
K=K'=\frac{\Gamma\left(\frac14\right)^{2}}{4\sqrt\pi}
\qquad E =E'= \frac{K}{2} +\frac{\pi}{4K}
\en

\vskip2ex
\noindent
{\bf G\&R 8.113-4, DLMF 19.5.1-2,19.12.1}
\eqa
\label{eq:EKtaylorseries}
k\rightarrow 0
\qquad
K(k) &\rightarrow \frac\pi2 \left(
1 + \frac{k^{2}}4 +\frac{9k^{4}}{64}  \right)+O(k^{6})
\\[1ex]
E(k) &\rightarrow \frac\pi2 \left(
1 - \frac{k^{2}}4  -\frac{3k^{4}}{64} \right)+O(k^{6}) 
\\[1ex]
K' &\rightarrow \ln \left(\frac4{k}\right)  + O(k^{2}\ln k)
\ena

\vskip2ex
\noindent
{\bf G\&R 8.123, DLMF 19.4(i)}
\eq
k\frac{dK}{dk} = \frac{E}{k'{}^{2}}-K
\qquad
k\frac{dE}{dk} = E-K
\en
which imply
\eq
k\frac{d}{dk} \frac{E}{K} = 
- \frac{1}{k'{}^{2}}\left(\frac{E}{K}\right)^{2}
+ 2 \frac{E}{K} -1
\en

\vskip2ex
\noindent
{\bf G\&R 8.159, DLMF 22.13.2}

\vskip1ex
The Jacobi elliptic function $\cn(z) = \cn(z,k)$ is the
analytic function (with poles) satisfying
\eq
\label{eq:cnidentity}
\cn'(z)^{2} = (1-\cn^{2})(k'{}^{2}+k^{2}\cn^{2})
\qquad
\cn(0) =1
\en
which is to say that $f(z) = k \cn(z,k)$ solves
\eq
\label{eq:kcneqn}
f'{}^{2} = (k^{2}-f^{2})(k'{}^{2}+f^{2})
\en

\vskip2ex
\noindent
{\bf G\&R 8.151, 8.146, DLMF 22.4(i), 22.5(ii)}

\vskip1ex
$\cn(z,k)$ is doubly periodic in $z$.
\eq
\cn(z) = \cn(z+4K) = \cn(z+4K'i) = \cn(z+2K +2K'i)
\en
On the real axis, $\cn(z,k)$ oscillates between $\pm 1$ with period 
$4K$.  In the limit $k\rightarrow 0$,
\eq
k\rightarrow 0 \qquad \cn(z,k) \rightarrow \cos\left(z\right)
\en

\vskip2ex
\noindent
{\bf G\&R 5.134, 5.131}

\eqa
\label{eq:cnintegrals}
\int_{0}^{4K} k^{2} \cn(z)^{2} \; dz  &= 4\left[ (k^{2}-1)K + E\right]
\\[2ex]
\int_{0}^{4K} k^{2} \cn'(z)^{2}\;du
&= \frac43 \left[(1-k^{2})K + (2k^{2}-1)E\right]
\ena

\end{appendix}

\phantomsection
\section*{Acknowledgments}
\addcontentsline{toc}{section}{\numberline{}Acknowledgments}
This work was supported by the Rutgers New High Energy Theory Center
and by the generosity of B. Weeks.
I am grateful
to the  Mathematics Division of the 
Science Institute of the University of Iceland
for its hospitality.

\bibliographystyle{ytphys}

\raggedright
\phantomsection
\bibliography{Literature}
\addcontentsline{toc}{section}{\numberline{}References}

\end{document}